\documentclass[11pt,a4paper]{article}
\usepackage[T1]{fontenc}
\usepackage[margin=30mm]{geometry}
\usepackage{mathtools}
\usepackage{amsthm}
\newtheorem{theorem}{Theorem}[section]
\newtheorem{lemma}[theorem]{Lemma}

\usepackage{booktabs}
\usepackage{adjustbox}
\usepackage{xurl}
\usepackage[numbers,square,sort&compress]{natbib}
\usepackage{caption}
\usepackage[section]{placeins}
\usepackage[hidelinks]{hyperref}
\usepackage{tikz}
\usetikzlibrary{arrows.meta,positioning,shapes.geometric,calc}
\tikzset{pidvalue/.style={font=\small,color=red!80!black,inner sep=1pt}}
\newcommand{\R}{\mathrm{R}}
\newcommand{\PR}{\mathrm{PR}}
\newcommand{\Ita}{\mathrm{I}}
\newcommand{\PI}{\mathrm{PI}}
\newcommand{\wt}{\operatorname{wt}}

\newenvironment{reductionfigure}{\begin{figure}[htbp]\centering}{\end{figure}}
\begin{document}

\title{Roman-Type Domination on Bisplit Graphs:\\Hardness and Inapproximability}
\author{M. S. Manjusha \quad P. Renjith\\[4pt]
  \small National Institute of Technology Calicut\\
  \small Kozhikode, Kerala, India\\
  \small \texttt{manjusha\_p220226cs@nitc.ac.in}, \texttt{renjith@nitc.ac.in}\\[3pt]
  \small Corresponding author: M. S. Manjusha}
\date{}
\maketitle

\begin{abstract}
Inspired by the structure and name of split graphs, bisplit graphs have
three independent sets $I,A,B$, where every vertex of $A$ is adjacent to
every vertex of $B$, and vertices of $I$ may have arbitrary neighborhoods
in the core $A\cup B$.  Thus they have a complete bipartite core together
with the remaining vertices, making them a
structured setting in which to study how core size affects complexity.
Mann and Fernau's construction already implies NP-completeness of perfect
Roman domination on bisplit graphs with a singleton core side
\citep[Theorem~20]{mann-fernau-pr}.  Motivated by this result, we study all
four Roman-type domination problems across further core-size regimes: we
prove NP-completeness when one core side has size one or two, and when both
sides have size at least three.
For perfect Italian domination, even deciding whether prescribed core
labels of total weight two can be extended is NP-complete on bipartite
bisplit graphs.
Approximation is also addressed: for every fixed $0<\varepsilon<1$,
unless $\mathrm{P}=\mathrm{NP}$, no $n^{1-\varepsilon}$-approximation
exists for perfect Roman domination when one core side has fixed size
$1$, $2$, or $3$, or for perfect Italian domination on bipartite bisplit
graphs whose two core sides grow with the input.  All four problems are
fixed-parameter tractable for both parameters.  Given supplied
modulators of size $k$, the running-time bases, in the order Roman,
perfect Roman, Italian, and perfect Italian, are $4,11,11,18$ for cluster
vertex deletion and $6,11,11,18$ for biclique-cluster vertex deletion,
up to polynomial factors.
The $4^k$ Roman cluster-deletion bound is known; the other bounds follow
from our dynamic programs.  To our knowledge, the latter parameter has
not previously been studied explicitly for domination variants.
\end{abstract}



\section{Introduction}\label{sec:prelim}

Roman-type domination studies minimum-weight labelings of graph vertices
by $0,1,2$, with constraints imposed only on vertices labeled $0$.  Roman
domination requires a neighbor labeled $2$; Italian domination allows
either such a neighbor or two neighbors labeled $1$.  Their perfect
variants require, respectively, exactly one neighbor labeled $2$ or total
neighbor-label sum exactly $2$.  These distinctions matter: adding a
positive label can violate an exact condition, so results do not
automatically transfer between variants
\citep{cockayne-roman,chellali-r2,banerjee-perfect-roman,haynes-henning-perfect-italian}.

For ordinary Roman domination, foundational work established the problem
and its basic bounds, while later work proved NP-hardness and developed
polynomial-time algorithms on structured classes, including linear-time
algorithms for interval graphs and cographs given their standard
representations~\citep{cockayne-roman,fernau-parameterized,liedloff-roman}.
Italian domination, also called Roman $\{2\}$-domination, has a related
incidence-based complexity literature: X3C reductions have been used to
claim NP-completeness on bisplit and other bipartite graph classes
\citep{padamutham-r2}.  Algorithmic results include linear-time
algorithms on chain and threshold graphs and a fixed-width treewidth
algorithm given a decomposition~\citep{padamutham-r2}.  As we explain
below, the reverse implication in the cited bisplit proof does not rule
out protection by two value-$1$ neighbors, and we supply a counterexample
and an independent hardness proof.

Perfect Roman domination is NP-complete on bipartite graphs, but has
linear-time algorithms on block graphs, cographs, and series--parallel
graphs~\citep{banerjee-perfect-roman}.  Perfect Italian domination is
NP-complete on chordal graphs and on bipartite planar graphs; linear-time
algorithms are known on block graphs, series--parallel graphs, and
cographs~\citep{pradhan-pi-complexity,lauri-mitillos,banerjee-pidf-cographs}.
Thus, even within the two perfect variants, complexity depends strongly
on the graph class.  \citet{pradhan-pi-complexity} also prove
an $n^{1-\varepsilon}$ inapproximability bound for perfect Italian
domination on general graphs.  Here we establish that bound on bipartite
bisplit graphs; the factor itself is not new.  Their Section~5.1
construction uses large cliques, whose edges are essential to the
forcing argument in Claim~9(b).  Those graphs are neither bipartite nor
bisplit.  Our contribution is the restricted graph class and the
prescribed-core extension result, rather than the gap-amplification
technique.

Bisplit graphs provide a structured setting between bipartite and split-like graph classes. A bisplit graph has a partition into stable sets $I,A,B$, with every
$A$--$B$ edge present and arbitrary edges from $I$ to $A\cup B$
\citep{brandstadt-bisplit}.  We call $A\cup B$ the core and $A,B$ its
sides.  \citet[Theorem~3]{padamutham-r2}
reported NP-completeness of Italian (Roman $\{2\}$-) domination on
bisplit graphs using a singleton core side.  Their reverse implication
overlooks the possibility of two value-1 neighbors.  In
Section~\ref{subsec:italian-singleton}, we give a counterexample to
that reduction and an independent proof of the claimed hardness.
We then examine how the constructions change with the core size.

Perfect Roman hardness with a singleton core side is already implicit
in \citet[Theorem~20]{mann-fernau-pr}.  In their notation, the
construction has the bisplit partition
$I=V'$, $A=\{x\}$, $B=V\cup Y\cup Z$.
Our exact-cover construction supplies an alternative proof shared
with Roman and perfect Italian domination, not a first hardness
result for this perfect Roman subclass.  The earlier bipartite
construction of \citet[Theorem~2]{banerjee-perfect-roman} and the
star-convex and comb-convex constructions of
\citet[Theorems~2.1--2.2]{padamutham-pr} are not bisplit.
Indeed, each has two nonadjacent neighbors of private leaves with
disjoint neighborhoods.  Both would have to lie on the same core
side, leaving no vertex for the nonempty opposite side.

The main contribution of this article is a systematic complexity map
for all four variants as the bisplit core changes size.  We establish
NP-completeness when one core side has size one or two, and when both
sides have size at least three; three of the variants are hard for every
fixed side size $t\geq2$.  The perfect-Roman singleton-side case was
already implicit in prior work; our results fill out the other core-size
regimes, correct the cited Italian-domination proof, and show that perfect
Italian domination remains NP-complete even when core labels of total
weight two are prescribed.  We also prove approximation lower bounds
on the stated bisplit subclasses and give fixed-parameter algorithms
for all four problems under vertex-cover, cluster-vertex-deletion, and
biclique-cluster-vertex-deletion parameters.  Roman domination parameterized
by cluster vertex deletion was previously studied in~\citep{ashok-rdf-cvd};
perfect Roman domination is also FPT by treewidth~\citep[Theorem~15]{mann-fernau-pr}.
The $n^{1-\varepsilon}$
lower bound for perfect Roman domination applies when one core side has
fixed size $1$, $2$, or $3$; for perfect Italian domination we establish
the bound when both core sides grow with the input.

Sections~\ref{sec:hardness}--\ref{sec:fixed-side-hardness} present
the hardness results by core structure: the first treats a singleton
side, while the second treats a side of size two, larger fixed sides, and
prescribed core labels for perfect Italian domination.  Section~\ref{sec:approx}
presents the approximation results, and Section~\ref{sec:parameterized}
gives the vertex-cover, cluster-vertex-deletion, and
biclique-cluster-vertex-deletion algorithms.

\subsection{Graph-theoretic preliminaries}\label{subsec:graph-prelim}

All graphs are finite, simple, and undirected.  Write $V(G)$ and $E(G)$
for the vertex and edge sets of a graph $G$, and put $n=|V(G)|$; thus,
the order of $G$ is $n$.
The size of $G$ (its number of edges) is $|E(G)|$.
The open neighborhood of a vertex $v$ is
$N(v)=\{u\in V(G):uv\in E(G)\}$, and its degree is $|N(v)|$.
A vertex of degree one is a leaf.  A set of vertices is stable, or
independent, if no two of its vertices are adjacent.  A vertex cover
is a set that contains at least one endpoint of every edge.
A graph is bipartite if its vertex set can be partitioned into two
stable sets; it is complete bipartite if every vertex in one part is
adjacent to every vertex in the other.

For $f:V(G)\to\{0,1,2\}$, put
$f(X)=\sum_{v\in X}f(v)$ and $\wt(f)=f(V(G))$.  The four rules are listed
in Table~\ref{tab:rules}; positive vertices impose no condition.
Their minimum weights are $\gamma_{\R}$, $\gamma_{\PR}$,
$\gamma_{\Ita}$, and $\gamma_{\PI}$.  The decision problems ask whether
these weights are at most a given integer $K$.  For each decision problem,
a certificate is a labeling $f:V(G)\to\{0,1,2\}$, which uses $O(n)$ bits.
A verifier sums the labels to check $\wt(f)\leq K$ and scans the edges to
check the required condition at every zero vertex: at least one or exactly
one neighbor labeled 2 for Roman or perfect Roman domination, respectively,
and neighbor-label sum at least or exactly 2 for Italian or perfect Italian
domination, respectively.  These checks take $O(n+|E(G)|)$ time, so all four
decision problems belong to NP.

\begin{table}[tb]
\centering
\caption{Conditions required at every vertex $v$ with $f(v)=0$}
\label{tab:rules}
\begin{tabular}{@{}ll@{}}
\toprule
Rule (function) & Condition \\
\midrule
Roman (RDF) & $|N(v)\cap f^{-1}(2)|\geq1$ \\
Perfect Roman (PRDF) & $|N(v)\cap f^{-1}(2)|=1$ \\
Italian (IDF) & $\sum_{u\in N(v)}f(u)\geq2$ \\
Perfect Italian (PIDF) & $\sum_{u\in N(v)}f(u)=2$ \\
\bottomrule
\end{tabular}
\end{table}

Throughout, the bisplit partition
$V(G)=I\mathbin{\dot\cup}A\mathbin{\dot\cup}B$ is supplied and both core
sides are nonempty.  Write $\alpha=f(A)$ and $\beta=f(B)$.
We use Set Cover and Exact Cover by 3-Sets (X3C): the former asks whether
at most $k$ sets cover a universe $U$; the latter asks whether a family
of triples contains a partition of a universe of size $3q$
\citep{garey-johnson}.  Restricted X3C (RXC3) additionally requires each
element to occur in exactly three sets and is NP-complete
\citep{gonzalez-x3c}.

\paragraph{Shared decision framework.}
Our decision reductions use an RXC3 instance $(U,\mathcal S)$, with
$U=\{e_1,\ldots,e_{3q}\}$ and $\mathcal S=\{S_1,\ldots,S_{3q}\}$.
We may assume $q\geq2$, since
smaller instances can be decided directly.  A cover by at most $q$
triples is necessarily exact.  We repeatedly use the following
forcing observation for all four rules: a family of more than $K$
vertices contains a zero vertex in every labeling of weight at most
$K$.  In particular, more than $K$ private leaves force their common
neighbor to have value 2.  Replicated element or test vertices then
impose the corresponding rule on their shared neighborhood.
The common concrete RXC3 example used in figures that display a specific
input is given below, with $q=2$.  Figures with generic indices or local
gadgets are schematic, not separate source instances.  The reductions
and proofs apply to arbitrary RXC3 instances:
\[
\begin{aligned}
\mathcal U&=\{e_1,e_2,e_3,e_4,e_5,e_6\},\\
\mathcal S&=\{S_1=\{e_1,e_2,e_3\},\quad
               S_2=\{e_1,e_2,e_4\},\\
             &\qquad S_3=\{e_1,e_5,e_6\},\quad
               S_4=\{e_3,e_5,e_6\},\\
             &\qquad S_5=\{e_2,e_4,e_5\},\quad
               S_6=\{e_3,e_4,e_6\}\}.
\end{aligned}
\]
Each element occurs in exactly three triples, and $\{S_2,S_4\}$ is an
exact cover of $\mathcal U$.  We first use this framework to prove
hardness when one core side is a singleton.

\section{Hardness with a Singleton Core Side}\label{sec:hardness}

We reduce from RXC3. A shared construction using $H_r$ proves hardness
for Roman, perfect Roman, and perfect Italian domination. The
Italian-domination proof uses a separate construction with additional
replicated tests to control value-1 labels.

\begin{theorem}\label{thm:all-singleton}
The decision problems for Roman, perfect Roman, Italian, and perfect Italian
domination are NP-complete on bisplit graphs supplied with a partition
satisfying $|A|=1$ and $|B|\geq3$.
\end{theorem}

\subsection{A shared construction for Roman and the perfect variants}
\label{subsec:common-construction}

\begin{figure}[!ht]\centering
\begin{adjustbox}{max width=\textwidth}
\begin{tikzpicture}[
v/.style={circle, fill=black, inner sep=0pt, minimum size=3.5pt},
vlabel/.style={font=\small, inner sep=2pt},
val/.style={font=\small, color=red!80!black, inner sep=2pt},
column_head/.style={font=\large\bfseries, align=center},
column_sub/.style={font=\small, align=center}
]

% --- Column Coordinates ---
\def\colD{0}
\def\colC{2.5}
\def\colX{5}
\def\colS{7.5}
\def\colU{11}

% --- Column Headers ---
\node[column_head] at (\colD, 5) {$I$};
\node[column_sub] at (\colD, 4.5) {forcing leaves};

\node[column_head] at (\colX, 5) {$I$};
\node[column_sub] at (\colX, 4.5) {Private test\\vertices};

\node[column_head] at (\colS, 5) {$B$};
\node[column_sub] at (\colS, 4.5) {set vertices};

\node[column_head] at (\colU, 5) {$I$};
\node[column_sub] at (\colU, 4.5) {element copies};

% =================================================================
% CENTER VERTEX c
% =================================================================
\node[v] (c) at (\colC, 0) {};
\node[vlabel, below=2pt of c] {$c$};
\node[val, below=15pt of c] {$2$};
\node[column_head] at (\colC, 4.8) {$A=\{c\}$};

% =================================================================
% FORCING LEAVES (d_1 to d_9)
% =================================================================
\foreach \i in {1,...,9} {
\pgfmathsetmacro{\y}{2.5 - 0.625*(\i-1)}
\node[v] (d\i) at (\colD, \y) {};
\node[vlabel, left=2pt of d\i] {$d_\i$};
\node[val, left=14pt of d\i] {$0$};
}

% =================================================================
% PRIVATE TEST VERTICES (x_1 to x_6)
% =================================================================
\node[v] (x1) at (\colX, 3.0) {};
\node[vlabel, above right=1pt of x1] {$x_1$};
\node[val, left=3pt of x1] {$0$};

\node[v] (x2) at (\colX, 2.1) {};
\node[vlabel, right=2pt of x2] {$x_2$};
\node[val, left=3pt of x2] {$0/1$};

\node[v] (x3) at (\colX, 1.2) {};
\node[vlabel, right=2pt of x3] {$x_3$};
\node[val, left=3pt of x3] {$0$};

\node[v] (x4) at (\colX, -2.1) {};
\node[vlabel, right=2pt of x4] {$x_4$};
\node[val, left=3pt of x4] {$0/1$};

\node[v] (x5) at (\colX, -3.0) {};
\node[vlabel, right=2pt of x5] {$x_5$};
\node[val, left=3pt of x5] {$0$};

\node[v] (x6) at (\colX, -3.9) {};
\node[vlabel, right=2pt of x6] {$x_6$};
\node[val, left=3pt of x6] {$0$};

% =================================================================
% SET VERTICES (S_1 to S_6)
% =================================================================
\node[v] (s1) at (\colS, 3.0) {};
\node[vlabel, right=2pt of s1] {$S_1$};
\node[val, left=3pt of s1] {$0$};

\node[v] (s2) at (\colS, 1.8) {};
\node[vlabel, right=2pt of s2] {$S_2$};
\node[val, left=3pt of s2] {$2$};

\node[v] (s3) at (\colS, 0.3) {};
\node[vlabel, right=2pt of s3] {$S_3$};
\node[val, left=3pt of s3] {$0$};

\node[v] (s4) at (\colS, -1.5) {};
\node[vlabel, right=2pt of s4] {$S_4$};
\node[val, left=3pt of s4] {$2$};

\node[v] (s5) at (\colS, -2.7) {};
\node[vlabel, right=2pt of s5] {$S_5$};
\node[val, left=3pt of s5] {$0$};

\node[v] (s6) at (\colS, -3.6) {};
\node[vlabel, right=2pt of s6] {$S_6$};
\node[val, left=3pt of s6] {$0$};

% =================================================================
% ELEMENT COPIES (u_1^1, u_1^9 to u_6^1, u_6^9)
% =================================================================
\foreach \i/\ytop/\ybot in {
1/3.8/3.1,
2/2.1/1.4,
3/0.4/-0.3,
4/-1.2/-1.9,
5/-2.4/-3.1,
6/-3.7/-4.4
} {
\node[v] (u\i_1) at (\colU, \ytop) {};
\node[vlabel, right=2pt of u\i_1] {$u_\i^1$};
\node[val, right=18pt of u\i_1] {$0$};

\node at (\colU, {(\ytop+\ybot)/2}) {$\vdots$};

\node[v] (u\i_9) at (\colU, \ybot) {};
\node[vlabel, right=2pt of u\i_9] {$u_\i^9$};
\node[val, right=18pt of u\i_9] {$0$};
}

% =================================================================
% EDGES
% =================================================================

% --- Red/Coral Edges: c to d_i ---
\begin{scope}[red!65!black, opacity=0.8]
\foreach \i in {1,...,9} {
\draw (c) -- (d\i);
}
\end{scope}

% --- Green Edges: c to x_i ---
\begin{scope}[draw=green!60!black, thick]
\foreach \i in {1,...,6} {
\draw (c) -- (x\i);
}
\end{scope}

% --- Blue Edges: c to S_j ---
\begin{scope}[draw=blue!75!black, thick]
\foreach \j in {1,...,6} {
\draw (c) -- (s\j);
}
\end{scope}

% --- Yellow Edges: x_j to S_j ---
\begin{scope}[yellow!85!black, thick]
\foreach \j in {1,...,6} {
\draw (x\j) -- (s\j);
}
\end{scope}

% --- Orange Edges: Set-Element Incidence Edges (S_j to u_i^1, u_i^9) ---
\begin{scope}[orange!85!black, opacity=0.85]
% S1 = {1, 2, 3}
\foreach \elem in {1, 2, 3} {
\draw (s1) -- (u\elem_1);
\draw (s1) -- (u\elem_9);
}

% S2 = {1, 2, 4}
\foreach \elem in {1, 2, 4} {
\draw (s2) -- (u\elem_1);
\draw (s2) -- (u\elem_9);
}

% S3 = {1, 5, 6}
\foreach \elem in {1, 5, 6} {
\draw (s3) -- (u\elem_1);
\draw (s3) -- (u\elem_9);
}

% S4 = {3, 5, 6}
\foreach \elem in {3, 5, 6} {
\draw (s4) -- (u\elem_1);
\draw (s4) -- (u\elem_9);
}

% S5 = {2, 4, 5}
\foreach \elem in {2, 4, 5} {
\draw (s5) -- (u\elem_1);
\draw (s5) -- (u\elem_9);
}

% S6 = {3, 4, 6}
\foreach \elem in {3, 4, 6} {
\draw (s6) -- (u\elem_1);
\draw (s6) -- (u\elem_9);
}
\end{scope}

\end{tikzpicture}
\end{adjustbox}
\caption{The common graph $H_r$ for Roman, perfect Roman, and perfect
Italian domination, illustrated using the RXC3 instance in
Section~\ref{sec:prelim} with $q=2$ and exact cover $\{S_2,S_4\}$.
Green edges join $c$ to the private-test vertices $x_j$, blue edges join
$c$ to the set vertices $S_j$, and yellow edges join each $x_j$ to $S_j$.
Orange edges encode the incidences $e_i\in S_j$, shown for the first and
$9$th copies of each element; red edges join $c$ to forcing leaves.
Vertical dots indicate omitted intermediate element copies.}
\label{fig:perfect-roman-construction}
\end{figure}
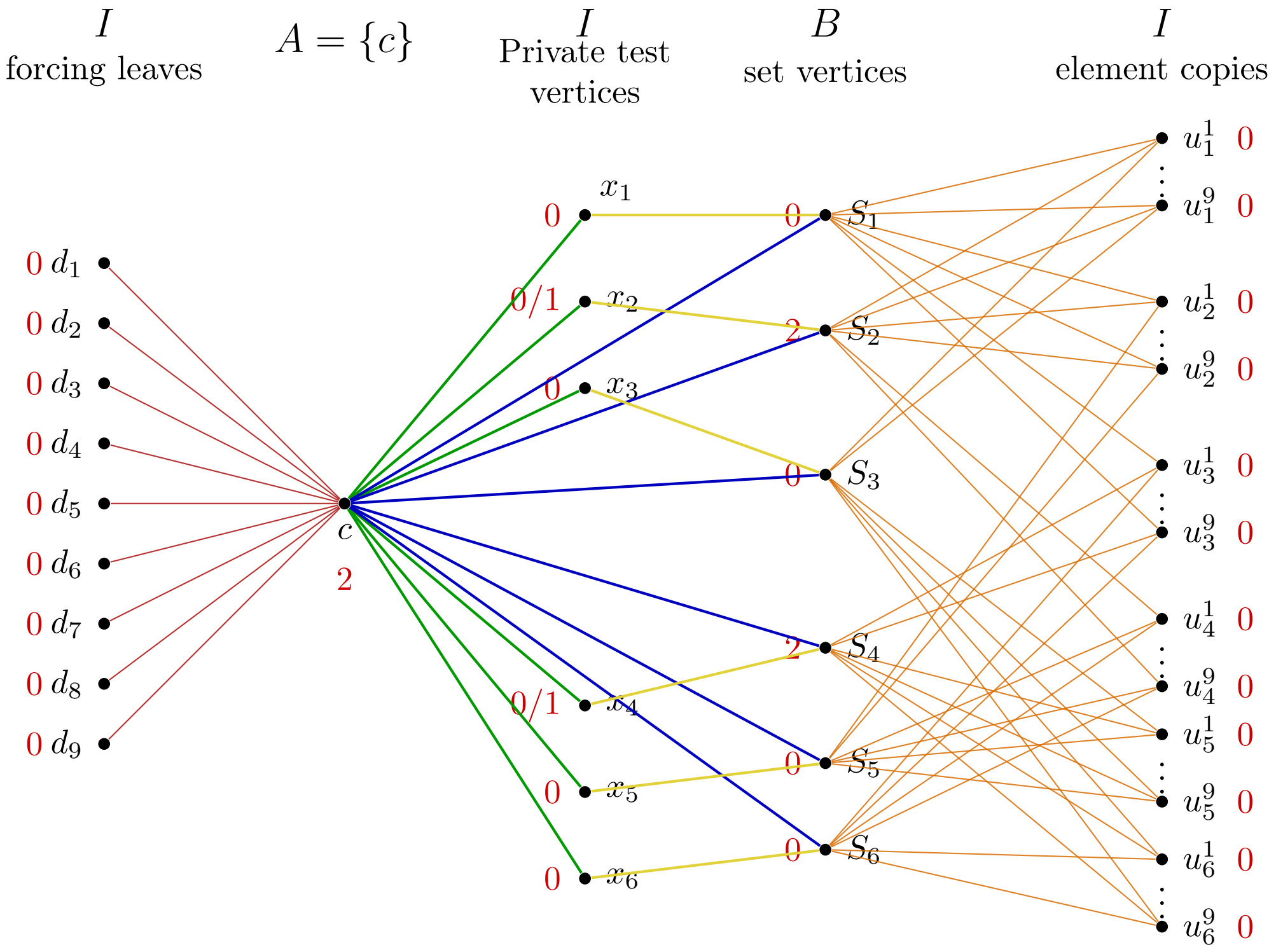

Given the RXC3 instance $(U,\mathcal S)$ and a replication parameter
$r$ (also used later for approximation hardness), construct the common graph $H_r$, illustrated in
Fig.~\ref{fig:perfect-roman-construction}, with
\[
\begin{aligned}
 A&=\{c\},\qquad B=\{s_j:1\leq j\leq 3q\},\\
 I&=\{d_\ell:1\leq\ell\leq r\}
 \mathbin{\dot\cup}\{x_j:1\leq j\leq 3q\}\\
 &\quad\mathbin{\dot\cup}
 \{u_i^\ell:1\leq i\leq3q,\ 1\leq\ell\leq r\}.
\end{aligned}
\]
Join $c$ to all of $B$.  The remaining edges are specified by
\[
\begin{aligned}
 N(d_\ell)&=\{c\},\qquad N(x_j)=\{c,s_j\},\\
 N(u_i^\ell)&=\{s_j:e_i\in S_j\}.
\end{aligned}
\]
There are no other edges.  This is a bisplit graph with
$1+6q+(3q+1)r$ vertices.  The decision reductions use $r=3q+3$ and
budgets
\[
 K_{\R}=2+2q,\qquad K_{\PR}=K_{\PI}=2+3q.
\]

\begin{proof}[Theorem~\ref{thm:all-singleton} for the three rules]
Given an exact cover, assign 2 to $c$ and its $q$ selected set
vertices.  For Roman domination, assign zero elsewhere.  For the
perfect variants, give each selected set's private test value 1 and
all remaining vertices value 0.  Every zero set vertex, private test,
or leaf is supported by $c$; each element copy is supported by its
unique selected set.  The weights are $K_{\R}$ and $K_{\PR}=K_{\PI}$,
respectively, and all required conditions hold.

Conversely, any labeling within its budget has $f(c)=2$ and a zero
copy of each element, because $r$ exceeds all three budgets.

\emph{Roman.} The value-2 set vertices cover $U$.  There are at most
$q$ of them, since their weight is at most $K_{\R}-2=2q$.
Covering $3q$ elements with at most $q$ triples requires an exact cover.

\emph{Perfect Roman.} Each zero element copy has exactly one value-2
set neighbor.  Thus those set vertices form an exact cover, consisting
of $q$ triples.  Their private tests must be positive because they see
both $c$ and the selected set with value 2.  These vertices use the
whole budget $2+3q$.

\emph{Perfect Italian.} A zero copy of each element gives
\begin{equation}\label{eq:singleton-element}
 \sum_{j:e_i\in S_j}f(s_j)=2.
\end{equation}
Summing over all elements, using that every set is a triple, yields
\begin{equation}\label{eq:singleton-total}
 \sum_{j=1}^3q f(s_j)=2q.
\end{equation}
Let $P=\{j:f(s_j)>0\}$.  For each $j\in P$, the private test $x_j$
must be positive: a zero test would receive $2+f(s_j)>2$.
Consequently,
\[
 K_{\PI}\geq\wt(f)\geq2+2q+|P|,\qquad |P|\geq q.
\]
Equality follows.  Exactly $q$ set vertices are positive, all with
value 2, so \eqref{eq:singleton-element} gives an exact cover.
All three constructions are polynomial, and membership in NP was
established in Section~\ref{sec:prelim}.
\end{proof}

\subsection{Italian domination}
\label{subsec:italian-singleton}

\paragraph{The earlier reduction.}
For an X3C instance $(X,\mathcal C)$ with $|X|=3q$ and
$\mathcal C=\{C_1,\ldots,C_t\}$, the construction in
\citet[Theorem~3]{padamutham-r2} has, in our notation,
\[
 I=X,\qquad A=\{a\},\qquad
 B=\{c_1,\ldots,c_t,a_1,a_2,a_3\}.
\]
The vertex $a$ is adjacent to all of $B$, and $x\in X$ is adjacent
to $c_j$ exactly when $x\in C_j$; there are no other edges.
Thus the core sides have sizes 1 and $t+3$.  The target budget is
$2q+2$.  The reverse proof assumes that a zero element vertex must
have a value-2 set neighbor, although two value-1 neighbors also
satisfy the Italian condition.

For a counterexample, take $X=\{1,\ldots,6\}$ and
\[
\begin{aligned}
 C_1&=\{1,2,3\}, & C_2&=\{1,4,5\},\\
 C_3&=\{2,4,6\}, & C_4&=\{3,5,6\}.
\end{aligned}
\]
Every two triples intersect, so no two form an exact cover.
Nevertheless, assigning 2 to $a$, 1 to all four $c_j$, and 0 to
every other vertex gives an IDF of weight $6=2q+2$.  Each element
occurs in exactly two triples and receives one unit from each;
each $a_i$ receives two from $a$.  Hence this NO-instance of X3C
maps to a YES-instance of Italian domination.  This disproves the
stated reduction equivalence, not the NP-completeness claim itself.

\paragraph{An independent reduction.}

\begin{proof}[Theorem~\ref{thm:all-singleton} for Italian domination]
Use the common RXC3 instance.  Put $K=10q+2$ and $R=K+1$, and define
\[
\begin{aligned}
 A&=\{c\},\\
 B&=\{s_j:1\leq j\leq 3q\}
       \mathbin{\dot\cup}\{x_i,y_i,z_i:1\leq i\leq3q\}.
\end{aligned}
\]
Add every $A$--$B$ edge.  The stable set $I$ consists of $R$ private
leaves $d_\ell$ of $c$ and, for each element $e_i$, four families
$\{u_{i,h}^\ell:1\leq\ell\leq R\}$, indexed by $h\in\{1,2,3,4\}$.
Their neighborhoods are
\begin{equation}\label{eq:italian-test-neighborhoods}
\begin{aligned}
 N(u_{i,1}^\ell)&=\{x_i,y_i\}, &
 N(u_{i,2}^\ell)&=\{x_i,z_i\},\\
 N(u_{i,3}^\ell)&=\{y_i,z_i\},\\
 N(u_{i,4}^\ell)&=\{x_i\}\cup\{s_j:e_i\in S_j\}.
\end{aligned}
\end{equation}
There are no other edges.  Thus $(I,A,B)$ is a bisplit partition with
$|A|=1$ and $|B|=12q\geq3$.  The graph has
$1+12q+R(1+12q)$ vertices and polynomially many edges.
Fig.~\ref{fig:italian-singleton-overview} gives an overview of the
construction.  Fig.~\ref{fig:italian-three-side} shows the local test gadgets.

\begin{reductionfigure}
\begin{adjustbox}{max width=\textwidth}
\begin{tikzpicture}[
    v/.style={circle, fill=black, inner sep=0pt, minimum size=3.5pt},
    vlabel/.style={font=\small, inner sep=2pt},
    val/.style={font=\small, color=red!80!black, inner sep=2pt},
    column_head/.style={font=\large\bfseries, align=center},
    column_sub/.style={font=\small, align=center}
]

  \def\colD{0}
  \def\colC{3.8}
  \def\colS{6.8}
  \def\colU{9.8}
  \def\colUFar{11.8}

  \node[column_head] at (\colD, 5.2) {$I$};
  \node[column_sub]  at (\colD, 4.6) {$R$ forcing leaves};
  \node[column_head] at (\colS, 5.2) {$B$};
  \node[column_sub]  at (\colS, 4.6) {Set vertices};
  \node[column_head] at (\colU, 5.2) {$I$};
  \node[column_sub]  at (\colU, 4.6) {four test families\\for $e_i$};

  \node[v] (c) at (\colC, -1.15) {};
  \node[vlabel, above=2pt of c] {$c$};
  \node[val, left=5pt of c, yshift=-8pt] {$2$};
  \node[column_head] at (\colC, 4.2) {$A=\{c\}$};

  \node[v] (d1) at (\colD, 1.6) {};
  \node[vlabel, left=2pt of d1] {$d_1$};
  \node[val, left=20pt of d1] {$0$};
  \node[v] (d2) at (\colD, 0.9) {};
  \node[vlabel, left=2pt of d2] {$d_2$};
  \node[val, left=20pt of d2] {$0$};
  \node[v] (d3) at (\colD, 0.2) {};
  \node[vlabel, left=2pt of d3] {$d_3$};
  \node[val, left=20pt of d3] {$0$};
  \node at (\colD, -0.8) {$\vdots$};
  \node[red] at (\colD+1.5, -0.8) {$\vdots$};
  \node[v] (d23) at (\colD, -1.8) {};
  \node[vlabel, left=2pt of d23] {$d_{23}$};
  \node[val, left=20pt of d23] {$0$};

  \node[v] (s1) at (\colS, 3.2) {};
  \node[vlabel, right=2pt of s1] {$S_1$};
  \node[val, left=5pt of s1] {$0$};
  \node[v] (s2) at (\colS, 2.4) {};
  \node[vlabel, right=2pt of s2] {$S_2$};
  \node[val, left=5pt of s2] {$1$};
  \node[v] (s3) at (\colS, 1.6) {};
  \node[vlabel, right=2pt of s3] {$S_3$};
  \node[val, left=5pt of s3] {$0$};
  \node[v] (s4) at (\colS, 0.8) {};
  \node[vlabel, right=2pt of s4] {$S_4$};
  \node[val, left=5pt of s4] {$1$};
  \node[v] (s5) at (\colS, 0.0) {};
  \node[vlabel, right=2pt of s5] {$S_5$};
  \node[val, left=5pt of s5] {$0$};
  \node[v] (s6) at (\colS, -0.8) {};
  \node[vlabel, right=2pt of s6] {$S_6$};
  \node[val, left=5pt of s6,yshift=3pt] {$0$};

  \node[v] (x1) at (\colS, -1.75) {};
  \node[vlabel, right=2pt of x1] {$x_1$};
  \node[val, left=5pt of x1,yshift=-3pt] {$1$};
  \node at (\colS, -2.1) {$\vdots$};
 \node[green!60!black] at (\colS-0.9, -1.75) {$\vdots$};
  \node[v] (x6) at (\colS, -2.6) {};
  \node[vlabel, right=2pt of x6,yshift=3pt] {$x_6$};
  \node[val, left=5pt of x6] {$1$};

  \node[v] (y1) at (\colS, -3.4) {};
  \node[vlabel, right=2pt of y1,yshift=3pt] {$y_1$};
  \node[val, left=5pt of y1] {$1$};
  \node at (\colS, -3.8) {$\vdots$};
  \node[green!60!black] at (\colS-0.9, -2.95) {$\vdots$};
  \node[v] (y6) at (\colS, -4.4) {};
  \node[vlabel, right=1pt of y6,yshift=3pt] {$y_6$};
  \node[val, left=5pt of y6,yshift=2pt] {$1$};

  \node[v] (z1) at (\colS, -5.1) {};
  \node[vlabel, right=0pt of z1,yshift=3pt] {$z_1$};
  \node[val, left=5pt of z1] {$1$};
  \node at (\colS, -5.5) {$\vdots$};
  \node[green!60!black] at (\colS-0.9, -4.1) {$\vdots$};
  \node[v] (z6) at (\colS, -6.1) {};
  \node[vlabel, right=1pt of z6,yshift=5pt] {$z_6$};
  \node[val, left=5pt of z6] {$1$};

  \node[v] (u1_4_1) at (\colU, 3.95) {};
  \node[vlabel, right=2pt of u1_4_1] {$u_{1,4}^1$};
  \node[val, right=22pt of u1_4_1] {$0$};
  \node at (\colU, 3.68) {$\vdots$};
  \node[v] (u1_4_23) at (\colU, 3.2) {};
  \node[vlabel, right=2pt of u1_4_23] {$u_{1,4}^{23}$};
  \node[val, right=22pt of u1_4_23] {$0$};
  \node[v] (u2_4_1) at (\colU, 2.7) {};
  \node[vlabel, right=2pt of u2_4_1] {$u_{2,4}^1$};
  \node[val, right=22pt of u2_4_1] {$0$};
  \node at (\colU, 2.4) {$\vdots$};
  \node[v] (u2_4_23) at (\colU, 1.9) {};
  \node[vlabel, right=2pt of u2_4_23] {$u_{2,4}^{23}$};
  \node[val, right=22pt of u2_4_23] {$0$};
  \node[v] (u3_4_1) at (\colU, 1.4) {};
  \node[vlabel, right=2pt of u3_4_1] {$u_{3,4}^1$};
  \node[val, right=22pt of u3_4_1] {$0$};
  \node at (\colU, 1.1) {$\vdots$};
  \node[v] (u3_4_23) at (\colU, 0.6) {};
  \node[vlabel, right=2pt of u3_4_23] {$u_{3,4}^{23}$};
  \node[val, right=22pt of u3_4_23] {$0$};
  \node[v] (u4_4_1) at (\colU, 0.1) {};
  \node[vlabel, right=2pt of u4_4_1] {$u_{4,4}^1$};
  \node[val, right=22pt of u4_4_1] {$0$};
  \node at (\colU, -0.09) {$\vdots$};
  \node[v] (u4_4_23) at (\colU, -0.55) {};
  \node[vlabel, right=2pt of u4_4_23] {$u_{4,4}^{23}$};
  \node[val, right=22pt of u4_4_23] {$0$};
  \node[v] (u5_4_1) at (\colU, -1.08) {};
  \node[vlabel, right=2pt of u5_4_1] {$u_{5,4}^1$};
  \node[val, right=22pt of u5_4_1] {$0$};
  \node at (\colU, -1.28) {$\vdots$};
  \node[v] (u5_4_23) at (\colU, -1.7) {};
  \node[vlabel, right=2pt of u5_4_23] {$u_{5,4}^{23}$};
  \node[val, right=22pt of u5_4_23] {$0$};
  \node[v] (u6_4_1) at (\colU, -2.2) {};
  \node[vlabel, right=2pt of u6_4_1] {$u_{6,4}^1$};
  \node[val, right=22pt of u6_4_1] {$0$};
  \node at (\colU, -2.4) {$\vdots$};
  \node[v] (u6_4_23) at (\colU, -2.8) {};
  \node[vlabel, right=2pt of u6_4_23] {$u_{6,4}^{23}$};
  \node[val, right=22pt of u6_4_23] {$0$};

  \node[v] (u1_1_1) at (\colU, -3.3) {};
  \node[vlabel, right=2pt of u1_1_1] {$u_{1,1}^1$};
  \node[val, right=22pt of u1_1_1] {$0$};
  \node at (\colU, -3.5) {$\vdots$};
  \node[v] (u1_1_23) at (\colU, -3.9) {};
  \node[vlabel, right=2pt of u1_1_23] {$u_{1,1}^{23}$};
  \node[val, right=22pt of u1_1_23] {$0$};
  \node[v] (u6_1_1) at (\colU, -4.5) {};
  \node[vlabel, right=2pt of u6_1_1] {$u_{6,1}^1$};
  \node[val, right=22pt of u6_1_1] {$0$};
  \node at (\colU, -4.7) {$\vdots$};
  \node[v] (u6_1_23) at (\colU, -5.1) {};
  \node[vlabel, right=2pt of u6_1_23] {$u_{6,1}^{23}$};
  \node[val, right=22pt of u6_1_23] {$0$};

  \node[v] (u1_3_1) at (\colU, -5.6) {};
  \node[vlabel, right=2pt of u1_3_1] {$u_{1,3}^1$};
  \node[val, right=22pt of u1_3_1] {$0$};
  \node at (\colU, -5.8) {$\vdots$};
  \node[v] (u1_3_23) at (\colU, -6.2) {};
  \node[vlabel, right=2pt of u1_3_23] {$u_{1,3}^{23}$};
  \node[val, right=22pt of u1_3_23] {$0$};
  \node[v] (u6_3_1) at (\colU, -6.9) {};
  \node[vlabel, right=2pt of u6_3_1] {$u_{6,3}^1$};
  \node[val, right=22pt of u6_3_1] {$0$};
  \node at (\colU, -7.1) {$\vdots$};
  \node[v] (u6_3_23) at (\colU, -7.6) {};
  \node[vlabel, right=2pt of u6_3_23] {$u_{6,3}^{23}$};
  \node[val, right=22pt of u6_3_23] {$0$};

  \node[v] (u1_2_1_far) at (\colUFar, -3.9) {};
  \node[vlabel, right=2pt of u1_2_1_far] {$u_{1,2}^1$};
  \node[val, right=22pt of u1_2_1_far] {$0$};
  \node at (\colUFar, -4.15) {$\vdots$};
  \node[v] (u1_2_23_far) at (\colUFar, -4.6) {};
  \node[vlabel, right=2pt of u1_2_23_far] {$u_{1,2}^{23}$};
  \node[val, right=22pt of u1_2_23_far] {$0$};
  \node[v] (u6_2_1_far) at (\colUFar, -5.45) {};
  \node[vlabel, right=2pt of u6_2_1_far] {$u_{6,2}^1$};
  \node[val, right=22pt of u6_2_1_far] {$0$};
  \node at (\colUFar, -5.65) {$\vdots$};
  \node[v] (u6_2_23_far) at (\colUFar, -6.1) {};
  \node[vlabel, right=2pt of u6_2_23_far] {$u_{6,2}^{23}$};
  \node[val, right=22pt of u6_2_23_far] {$0$};

  \begin{scope}[red!65!black, opacity=0.8]
    \draw (c) -- (d1);
    \draw (c) -- (d2);
    \draw (c) -- (d3);
    \draw (c) -- (d23);
  \end{scope}

  \begin{scope}[green!60!black, thick]
    \draw (c) -- (s1); \draw (c) -- (s2); \draw (c) -- (s3);
    \draw (c) -- (s4); \draw (c) -- (s5); \draw (c) -- (s6);
    \draw (c) -- (x1); \draw (c) -- (x6);
    \draw (c) -- (y1); \draw (c) -- (y6);
    \draw (c) -- (z1); \draw (c) -- (z6);
  \end{scope}

  \begin{scope}[orange!85!black, opacity=0.85]
    \foreach \u in {u1_4_1, u1_4_23, u2_4_1, u2_4_23, u3_4_1, u3_4_23} { \draw (s1) -- (\u); }
    \foreach \u in {u1_4_1, u1_4_23, u2_4_1, u2_4_23, u4_4_1, u4_4_23} { \draw (s2) -- (\u); }
    \foreach \u in {u1_4_1, u1_4_23, u5_4_1, u5_4_23, u6_4_1, u6_4_23} { \draw (s3) -- (\u); }
    \foreach \u in {u3_4_1, u3_4_23, u5_4_1, u5_4_23, u6_4_1, u6_4_23} { \draw (s4) -- (\u); }
    \foreach \u in {u2_4_1, u2_4_23, u4_4_1, u4_4_23, u5_4_1, u5_4_23} { \draw (s5) -- (\u); }
    \foreach \u in {u3_4_1, u3_4_23, u4_4_1, u4_4_23, u6_4_1, u6_4_23} { \draw (s6) -- (\u); }
  \end{scope}

  \draw[cyan!60!white, thin] (x1) -- (u1_4_1);
  \draw[cyan!60!white, thin] (x1) -- (u1_4_23);
  \draw[cyan!60!white, thin] (x6) -- (u6_4_1);
  \draw[cyan!60!white, thin] (x6) -- (u6_4_23);

  \begin{scope}[blue!80!black, thick]
    \draw (x1) -- (u1_1_1); \draw (x1) -- (u1_1_23);
    \draw (y1) -- (u1_1_1); \draw (y1) -- (u1_1_23);
    \draw (x6) -- (u6_1_1); \draw (x6) -- (u6_1_23);
    \draw (y6) -- (u6_1_1); \draw (y6) -- (u6_1_23);
  \end{scope}

  \begin{scope}[purple!80!black, thick]
    \draw (y1) -- (u1_3_1); \draw (y1) -- (u1_3_23);
    \draw (y6) -- (u6_3_1); \draw (y6) -- (u6_3_23);
    \draw (z1) -- (u1_3_1); \draw (z1) -- (u1_3_23);
    \draw (z6) -- (u6_3_1); \draw (z6) -- (u6_3_23);
  \end{scope}

  \begin{scope}[yellow!85!black, thick]
    \draw (x1) -- (u1_2_1_far); \draw (x1) -- (u1_2_23_far);
    \draw (x6) -- (u6_2_1_far); \draw (x6) -- (u6_2_23_far);
    \draw (z1) -- (u1_2_1_far); \draw (z1) -- (u1_2_23_far);
    \draw (z6) -- (u6_2_1_far); \draw (z6) -- (u6_2_23_far);
  \end{scope}

\end{tikzpicture}
\end{adjustbox}
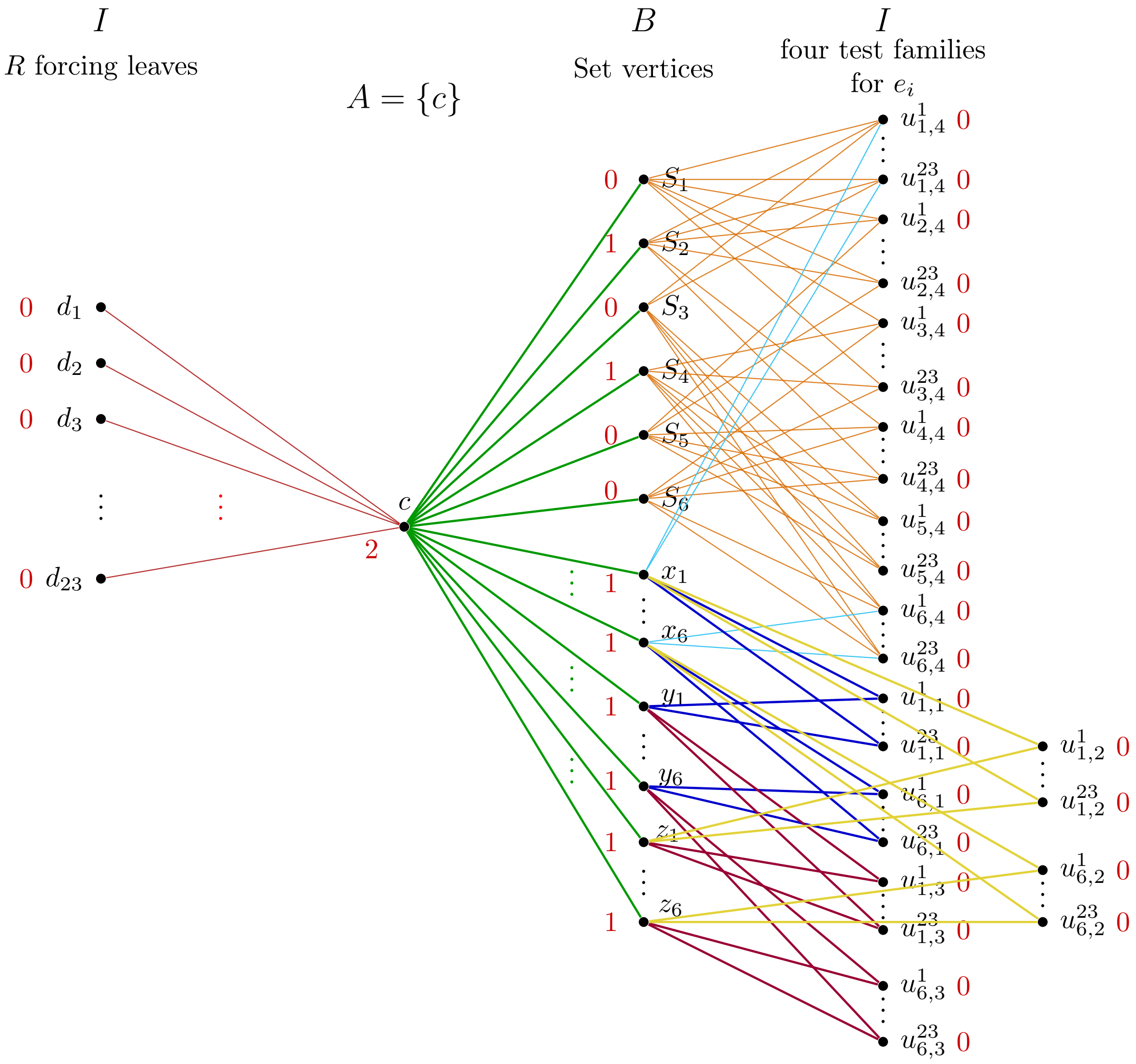
\caption{Overview of the singleton-side Italian construction, illustrated
for $q=2$ and $R=23$ using the exact cover $\{S_2,S_4\}$. Blue edges
represent the complete bipartite core between $A$ and $B$. The red values
give the corresponding IDF: $c$ has value 2, every auxiliary vertex and
each selected set vertex has value 1, and all remaining vertices have
value 0. Only representative vertices and edges are shown; dotted gaps
indicate omitted indexed vertices and copies.}
\label{fig:italian-singleton-overview}
\vspace{2mm}
\begin{adjustbox}{max width=\textwidth}
\begin{tikzpicture}[
 v/.style={circle,fill=black,inner sep=0pt,minimum size=5pt},
 lab/.style={font=\small,align=center}]
 \node[v,label=above:{$x_i$}] (x) at (0,1.3) {};
 \node[v,label=above:{$y_i$}] (y) at (1.8,1.3) {};
 \node[v,label=above:{$z_i$}] (z) at (3.6,1.3) {};
 \node[v,label=below:{$u_{i,1}^\ell$}] (u1) at (0,-0.4) {};
 \node[v,label=below:{$u_{i,2}^\ell$}] (u2) at (1.8,-0.4) {};
 \node[v,label=below:{$u_{i,3}^\ell$}] (u3) at (3.6,-0.4) {};
 \draw (u1)--(x) (u1)--(y) (u2)--(x) (u2)--(z)
       (u3)--(y) (u3)--(z);
 \foreach \v in {x,y,z}
   \node[pidvalue,above=17pt of \v] {$1$};
 \foreach \v in {u1,u2,u3}
   \node[pidvalue,below=20pt of \v] {$0$};
 \node[lab] at (1.8,2.5) {(a) pair tests};
 \node[v,label=above:{$x_i$}] (xx) at (6,1.3) {};
 \node[v,label=above:{$s_j$}] (sj) at (7.5,1.3) {};
 \node[v,label=above:{$s_k$}] (sk) at (9,1.3) {};
 \node[v,label=above:{$s_\ell$}] (sl) at (10.5,1.3) {};
 \node[v,label=below:{$u_{i,4}^\ell$}] (u4) at (8.25,-0.4) {};
 \draw (u4)--(xx) (u4)--(sj) (u4)--(sk) (u4)--(sl);
 \node[pidvalue,above=17pt of xx] {$1$};
 \node[pidvalue,above=17pt of sj] {$1$};
 \node[pidvalue,above=17pt of sk] {$0$};
 \node[pidvalue,above=17pt of sl] {$0$};
 \node[pidvalue,below=20pt of u4] {$0$};
 \node[lab] at (8.7,2.5) {(b) element test};
\end{tikzpicture}
\end{adjustbox}
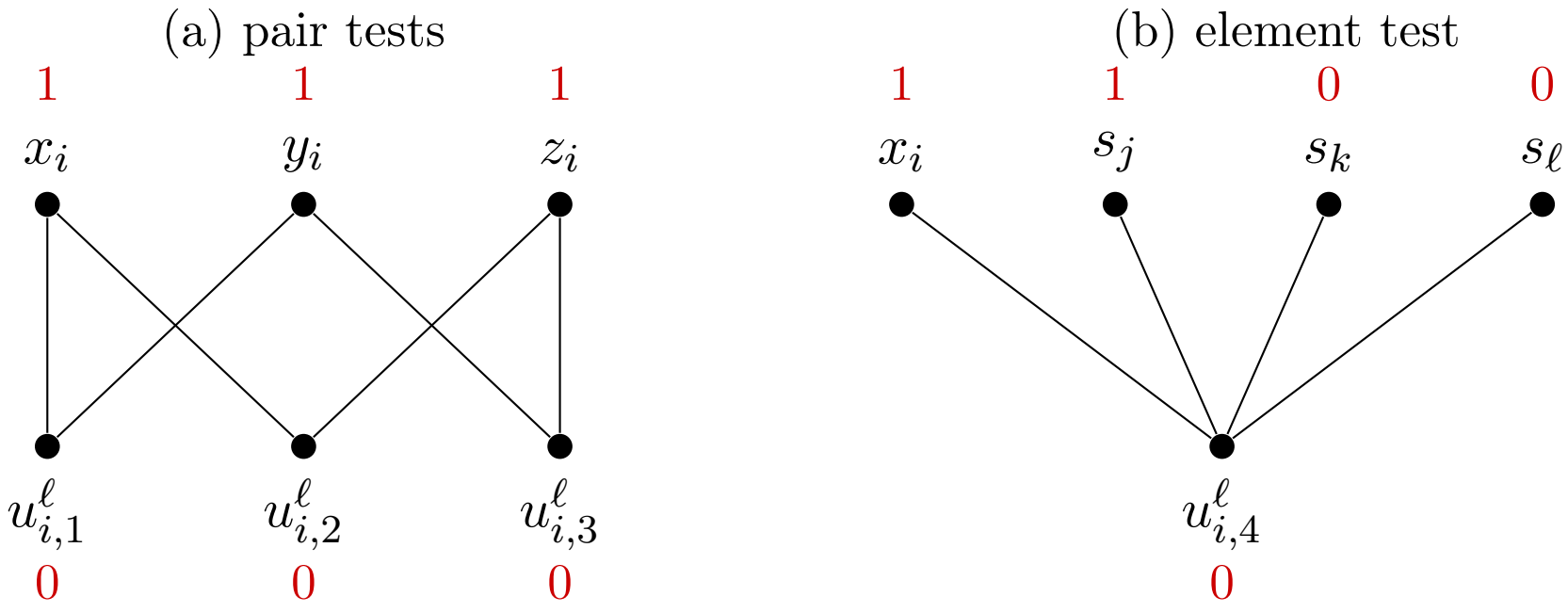
\caption{Local IDF gadgets for the Italian construction. Each test
represents $R$ copies. In (a), the upper vertices lie in $B$ and the
tests in $I$. In (b), the three sets containing $e_i$ are shown: $S_j$ is selected,
while $S_k$ and $S_\ell$ are not.}
\label{fig:italian-three-side}
\end{reductionfigure}

Suppose an exact cover exists.  Assign 2 to $c$, 1 to every
$x_i,y_i,z_i$, and 1 to the $q$ selected set vertices.  Assign zero
elsewhere.  Every pair-test vertex receives two units.  Every
element-test vertex receives one unit from $x_i$ and one from its
unique selected set.  The forcing leaves and all zero vertices in
$B$ receive two from $c$.  This is an IDF of weight
$2+9q+q=K$.

Conversely, let $f$ be an IDF of weight at most $K$.  The $R>K$
private leaves force $f(c)=2$.  Each test family has $R=K+1$
copies, more than the budget $K$; hence each has a zero copy, so
\eqref{eq:italian-test-neighborhoods} gives
\begin{equation}\label{eq:italian-test-constraints}
\begin{aligned}
 f(x_i)+f(y_i)&\geq2,\\
 f(x_i)+f(z_i)&\geq2,\\
 f(y_i)+f(z_i)&\geq2,\\
 f(x_i)+\sum_{j:e_i\in S_j}f(s_j)&\geq2.
\end{aligned}
\end{equation}
Let $\mathcal S_f=\{S_j:f(s_j)>0\}$, and let $s$ be the number
of elements not covered by $\mathcal S_f$.  Adding the first three
inequalities gives
$2(f(x_i)+f(y_i)+f(z_i))\geq6$, so the three auxiliary labels for
each element have total weight at least 3.  This is their baseline
cost.  If $e_i$ is uncovered, every incident set vertex has label 0,
and the fourth inequality forces $f(x_i)\geq2$.  The third inequality
also gives $f(y_i)+f(z_i)\geq2$, so the auxiliary labels for $e_i$
total at least 4, one above the baseline.

Define the set-vertex weight $W$ and the excess auxiliary weight $D$ by
\[
\begin{aligned}
 W&=\sum_{j=1}^{3q} f(s_j),\\
 D&=\sum_{i=1}^{3q}\bigl(f(x_i)+f(y_i)+f(z_i)-3\bigr)\geq0.
\end{aligned}
\]
Since $f(c)=2$, the total weight is at least $2+9q+W+D$.
The budget $K=10q+2$ therefore leaves at most $q$ units to pay for
set labels and auxiliary weight above the baseline, giving $W+D\leq q$.
Each uncovered element contributes at least one to $D$, so $s\leq D$.
Also
$|\mathcal S_f|\leq W$, so the triples in $\mathcal S_f$ cover at
most $3W$ elements.  Therefore
\[
\begin{aligned}
 3q&\leq3W+s\leq3W+D\\
   &=3(W+D)-2D\leq3q-2D.
\end{aligned}
\]
It follows that $D=0$, $W=q$, and $s=0$.  Thus at most $q$ triples
cover all $3q$ elements.  They must be exactly $q$ pairwise disjoint
triples, giving an exact cover.  The reduction is polynomial, and
Italian domination belongs to NP.
\end{proof}

Thus all four domination problems are NP-complete when one core side is a singleton.
Section~\ref{sec:fixed-side-hardness} treats a fixed-size core side and prescribed core labels.

\section{Hardness with a Fixed-Size Core Side and Prescribed Core Labels}
\label{sec:fixed-side-hardness}

First, we extend the Roman, perfect Roman, and Italian reductions to every
fixed integer size $t\geq2$ of one core side.  We also prove perfect Italian
hardness when that side has size two.  The final subsection establishes
NP-completeness of deciding whether prescribed core labels extend to a
PIDF, even when both core sides have size at least three.

\begin{theorem}\label{thm:all-two-side}
The decision problems for Roman, perfect Roman, Italian, and perfect Italian
domination are NP-complete on bisplit graphs supplied with a partition
satisfying $|A|=2$ and $|B|\geq3$.
\end{theorem}

\subsection{Hardness for Roman, perfect Roman, and Italian domination}

\begin{lemma}\label{lem:core-extension}
For every fixed integer $t\geq2$, the decision problems for Roman,
perfect Roman, and Italian domination are NP-complete on bisplit graphs
supplied with a partition satisfying $|A|=t$ and $|B|\geq3$.
\end{lemma}

\begin{proof}
In each construction, add $t-1$ vertices $a_1,\ldots,a_{t-1}$ to $A$,
each adjacent exactly to $B$.  Thus $A=\{c,a_1,\ldots,a_{t-1}\}$;
all other edges remain as before.
Fig.~\ref{fig:roman-core-extensions} shows the Roman, perfect Roman,
and Italian core-extension label patterns.

\begin{reductionfigure}
\begin{adjustbox}{max width=\textwidth}
\begin{tikzpicture}[
 v/.style={circle,fill=black,inner sep=0pt,minimum size=5pt},
 lab/.style={font=\small,align=center},
 every label/.style={font=\small}]
 \foreach \shift/\aval/\heading in {0/0/{(a) Roman: RDF},6.5/1/{(b) perfect Roman: PRDF}}{
  \begin{scope}[xshift=\shift cm]
   \node[v,label=left:{$c$}] (c) at (0,1.2) {};
   \node[v,label=left:{$a_1$}] (a1) at (0,0) {};
   \node[v,label=left:{$a_{t-1}$}] (aLast) at (0,-1.2) {};
   \node[lab] at (0,-0.6) {$\vdots$};
   \node[v,label=right:{$s_1$}] (sj) at (3,1.2) {};
   \node[v,label=right:{$s_2$}] (sk) at (3,0) {};
   \node[v,label=right:{$s_{3q}$}] (sh) at (3,-1.2) {};
   \node[lab] at (3,-0.6) {$\vdots$};
   \foreach \s in {sj,sk,sh}{
    \draw[blue!55] (c)--(\s) (a1)--(\s);
    \draw[blue!55,dashed] (aLast)--(\s);
   }
   \node[pidvalue,left=20pt of c] {$2$};
   \node[pidvalue,left=27pt of a1] {$\aval$};
   \node[pidvalue,left=27pt of aLast] {$\aval$};
   \node[pidvalue,right=24pt of sj] {$0/2$};
   \node[pidvalue,right=24pt of sk] {$0/2$};
   \node[pidvalue,right=24pt of sh] {$0/2$};
   \node[lab] at (0,2) {$A$};
   \node[lab] at (3,2) {$B$};
   \node[lab] at (1.5,2.8) {\heading};
  \end{scope}
 }
 \begin{scope}[xshift=2.75cm,yshift=-1.5cm,scale=0.85,transform shape]
  \node[v,label=left:{$c$}] (c) at (1,-2.7) {};
  \node[v,label=left:{$a_1$}] (a1) at (1,-3.7) {};
  \node[v,label=left:{$a_{t-1}$}] (aLast) at (1,-4.7) {};
  \node[lab] at (1,-4.2) {$\vdots$};
  \node[v,label=right:{$s_j$}] (ss) at (5,-2.3) {};
  \node[v,label=right:{$x_i$}] (xb) at (5,-3.1) {};
  \node[v,label=right:{$y_i$}] (yb) at (5,-3.9) {};
  \node[v,label=right:{$z_i$}] (zb) at (5,-4.7) {};
  \node[v,label=left:{$d_\ell$}] (d) at (-1,-2.7) {};
  \foreach \b in {ss,xb,yb,zb}{
   \draw[blue!55] (c)--(\b) (a1)--(\b);
   \draw[blue!55,dashed] (aLast)--(\b);
  }
  \draw (c)--(d);
  \node[pidvalue,above=5pt of c] {$2$};
  \node[pidvalue,left=27pt of a1] {$0$};
  \node[pidvalue,left=27pt of aLast] {$0$};
  \node[pidvalue,above=5pt of d] {$0$};
  \node[pidvalue,right=27pt of ss] {$0/1$};
  \node[pidvalue,right=27pt of xb] {$1$};
  \node[pidvalue,right=27pt of yb] {$1$};
  \node[pidvalue,right=27pt of zb] {$1$};
  \node[lab] at (2.5,-1.5) {(c) Italian: IDF};
 \end{scope}
\end{tikzpicture}
\end{adjustbox}
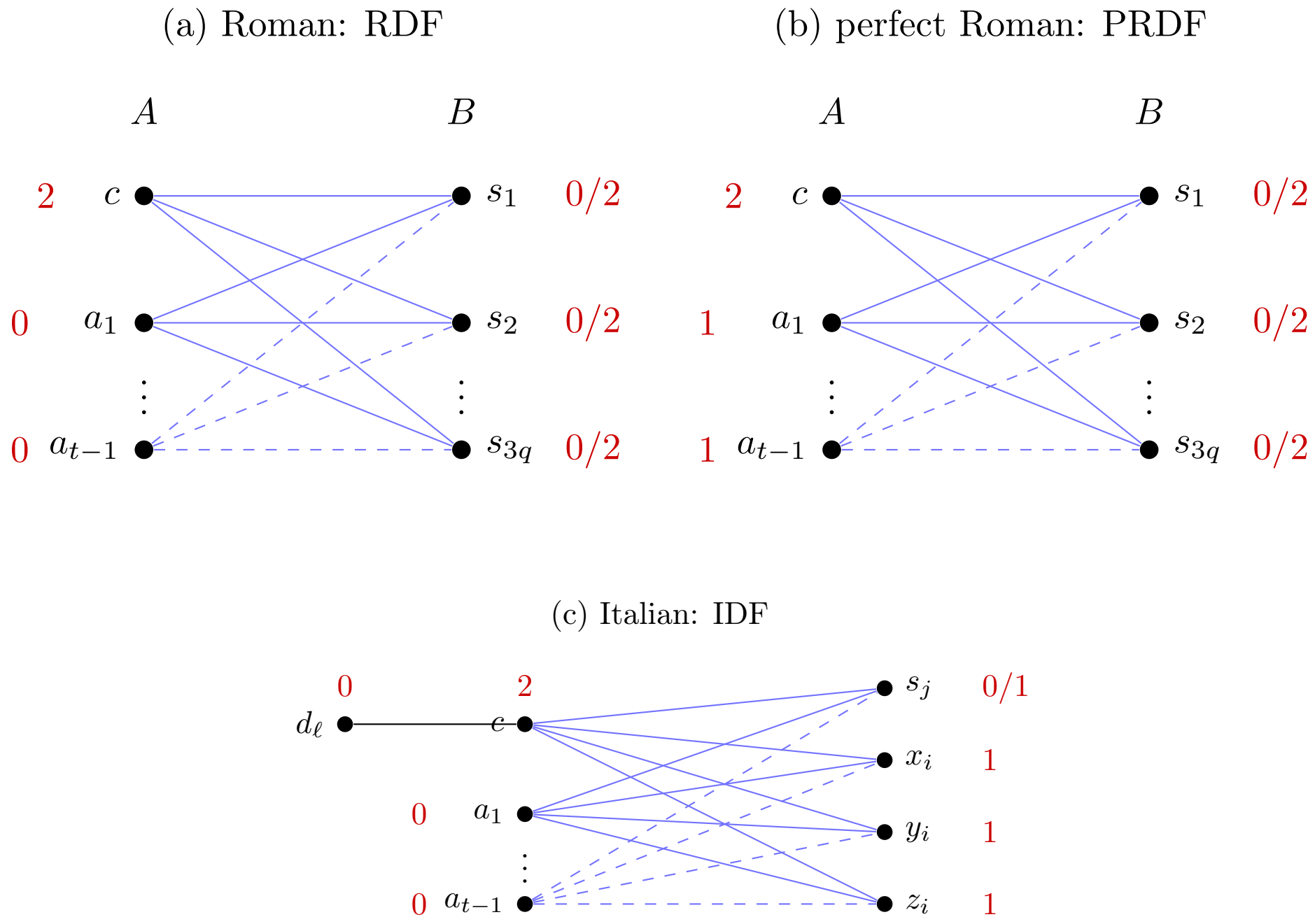
\caption{Core-extension labelings: (a) Roman, (b) perfect Roman, and
(c) Italian domination. In panels (a,b), each set vertex $s_1,s_2,\ldots,s_{3q}$ may be labeled 0 or 2 (shown as $0/2$). Panel (c) shows $s_j,x_i,y_i,z_i$, with $1\le i,j\le3q$: $s_j$ indexes
sets, and $x_i,y_i,z_i$ correspond to $e_i$. The $A$-side shows
$c,a_1,\ldots,a_{t-1}$ (only $a_1$ if $t=2$). Red numbers are labels.}
\label{fig:roman-core-extensions}
\end{reductionfigure}

For Roman domination, use the common graph $H_r$ from
Section~\ref{subsec:common-construction}, with $K=2+2q$ and
$r=K+1$. An exact cover gives a labeling of weight $K$: label $c$ and
the selected set vertices 2, and all other vertices, including every
added $a_i$, 0. Each added $a_i$ is adjacent to all $q$ selected set vertices labeled 2, so it is dominated. Conversely, since $r>K$, the private leaves force $f(c)=2$
and the replicated element copies force every element to be covered by
a value-2 set vertex. The added $a_i$ vertices cannot replace these set
choices because they are not adjacent to element copies. The budget then
allows at most $q$ selected sets, so they form an exact cover.

For perfect Roman domination, use the same common graph $H_r$ from
Section~\ref{subsec:common-construction}, with $K=2+3q+(t-1)$ and
$r=K+1$, retaining all private tests.  An exact cover gives the original
labeling with each $a_i$ labeled 1, of weight $K$; these new labels do
not change any old vertex's number of value-2 neighbors.  Conversely,
the leaves force $f(c)=2$ and zero element copies force an exact cover.
Its $q$ selected sets force their private tests and every $a_i$ to be
positive, using the entire budget $2+3q+(t-1)$.

For Italian domination, retain $K=10q+2$ and the graph of
Section~\ref{subsec:italian-singleton}.  An exact cover extends with
$f(a_i)=0$, since $f(B)=10q\geq2$.  The leaves and test neighborhoods
are unchanged, so the reverse proof still gives
\eqref{eq:italian-test-constraints} and $W+D\leq q$, and hence an exact
cover.  Each construction is polynomial, and all three problems are in NP.
\end{proof}

\subsection{Perfect Italian domination with a two-vertex core side}

The new core vertices must supply two units together.  We replace the
center by two vertices with the same neighbors and give each value one.

\begin{lemma}\label{lem:pi-two-side}
The decision problem for perfect Italian domination is NP-complete on
bisplit graphs supplied with a partition satisfying $|A|=2$ and $|B|\geq3$.
\end{lemma}

\begin{proof}
Use the common RXC3 instance and $H_r$ from
Section~\ref{subsec:common-construction}, with $K=2+3q$ and $r=K+1$.
Add a vertex
$c'$ to $A$ with exactly the same neighbors as $c$, but with no edge
between $c$ and $c'$.  Thus
\[
 A=\{c,c'\},\qquad B=\{s_1,\ldots,s_{3q}\},
\]
and the independent-set neighborhoods are
\[
\begin{aligned}
 N(d_\ell)&=\{c,c'\},\\
 N(x_j)&=\{c,c',s_j\},\\
 N(u_i^\ell)&=\{s_j:e_i\in S_j\}.
\end{aligned}
\]
Every $A$--$B$ edge is present, and there are no other edges.
The vertices $d_\ell$ now have degree two, not degree one.
The construction is polynomial; its labeling is illustrated in
Fig.~\ref{fig:pi-two-side}.

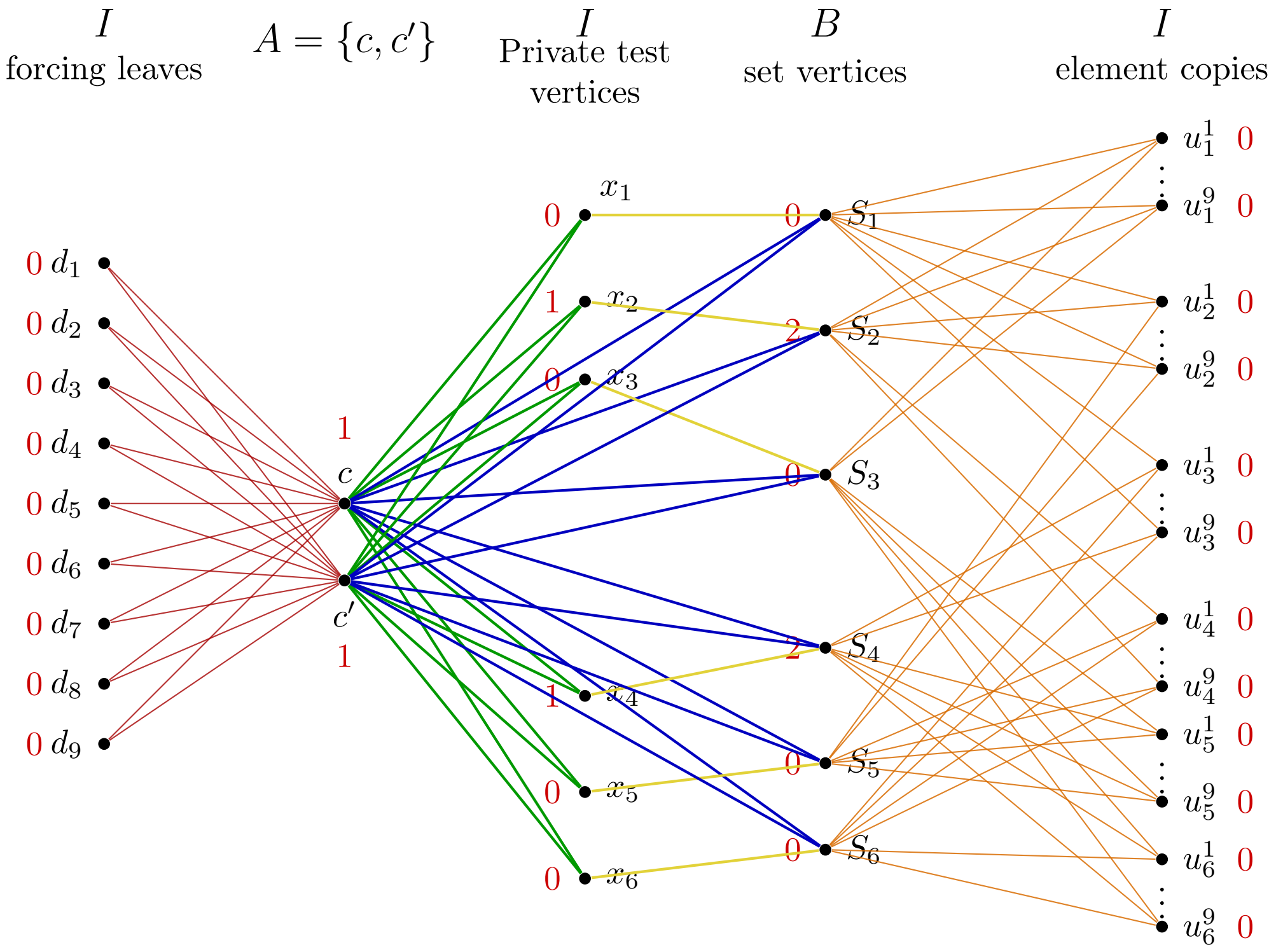
\begin{figure}[!ht]\centering
\begin{adjustbox}{max width=\textwidth}
\begin{tikzpicture}[
v/.style={circle, fill=black, inner sep=0pt, minimum size=3.5pt},
vlabel/.style={font=\small, inner sep=2pt},
val/.style={font=\small, color=red!80!black, inner sep=2pt},
column_head/.style={font=\large\bfseries, align=center},
column_sub/.style={font=\small, align=center}
]

% --- Column Coordinates ---
\def\colD{0}
\def\colC{2.5}
\def\colX{5}
\def\colS{7.5}
\def\colU{11}

% --- Column Headers ---
\node[column_head] at (\colD, 5) {$I$};
\node[column_sub] at (\colD, 4.5) {forcing leaves};

\node[column_head] at (\colX, 5) {$I$};
\node[column_sub] at (\colX, 4.5) {Private test\\vertices};

\node[column_head] at (\colS, 5) {$B$};
\node[column_sub] at (\colS, 4.5) {set vertices};

\node[column_head] at (\colU, 5) {$I$};
\node[column_sub] at (\colU, 4.5) {element copies};

% =================================================================
% CENTER VERTEX c
% =================================================================
\node[v] (c) at (\colC, 0) {};
\node[vlabel, above=2pt of c] {$c$};
\node[val, above=15pt of c] {$1$};
\node[v] (cp) at (\colC, -0.8) {};
\node[vlabel, below=2pt of cp] {$c'$};
\node[val, below=15pt of cp] {$1$};
\node[column_head] at (\colC, 4.8) {$A=\{c,c'\}$};

% =================================================================
% FORCING LEAVES (d_1 to d_9)
% =================================================================
\foreach \i in {1,...,9} {
\pgfmathsetmacro{\y}{2.5 - 0.625*(\i-1)}
\node[v] (d\i) at (\colD, \y) {};
\node[vlabel, left=2pt of d\i] {$d_\i$};
\node[val, left=14pt of d\i] {$0$};
}

% =================================================================
% PRIVATE TEST VERTICES (x_1 to x_6)
% =================================================================
\node[v] (x1) at (\colX, 3.0) {};
\node[vlabel, above right=1pt of x1] {$x_1$};
\node[val, left=3pt of x1] {$0$};

\node[v] (x2) at (\colX, 2.1) {};
\node[vlabel, right=2pt of x2] {$x_2$};
\node[val, left=3pt of x2] {$1$};

\node[v] (x3) at (\colX, 1.29) {};
\node[vlabel, right=2pt of x3] {$x_3$};
\node[val, left=3pt of x3] {$0$};

\node[v] (x4) at (\colX, -2) {};
\node[vlabel, right=2pt of x4] {$x_4$};
\node[val, left=3pt of x4] {$1$};

\node[v] (x5) at (\colX, -3.0) {};
\node[vlabel, right=2pt of x5] {$x_5$};
\node[val, left=3pt of x5] {$0$};

\node[v] (x6) at (\colX, -3.9) {};
\node[vlabel, right=2pt of x6] {$x_6$};
\node[val, left=3pt of x6] {$0$};

% =================================================================
% SET VERTICES (S_1 to S_6)
% =================================================================
\node[v] (s1) at (\colS, 3.0) {};
\node[vlabel, right=2pt of s1] {$S_1$};
\node[val, left=3pt of s1] {$0$};

\node[v] (s2) at (\colS, 1.8) {};
\node[vlabel, right=2pt of s2] {$S_2$};
\node[val, left=3pt of s2] {$2$};

\node[v] (s3) at (\colS, 0.3) {};
\node[vlabel, right=2pt of s3] {$S_3$};
\node[val, left=3pt of s3] {$0$};

\node[v] (s4) at (\colS, -1.5) {};
\node[vlabel, right=2pt of s4] {$S_4$};
\node[val, left=3pt of s4] {$2$};

\node[v] (s5) at (\colS, -2.7) {};
\node[vlabel, right=2pt of s5] {$S_5$};
\node[val, left=3pt of s5] {$0$};

\node[v] (s6) at (\colS, -3.6) {};
\node[vlabel, right=2pt of s6] {$S_6$};
\node[val, left=3pt of s6] {$0$};

% =================================================================
% ELEMENT COPIES (u_1^1, u_1^9 to u_6^1, u_6^9)
% =================================================================
\foreach \i/\ytop/\ybot in {
1/3.8/3.1,
2/2.1/1.4,
3/0.4/-0.3,
4/-1.2/-1.9,
5/-2.4/-3.1,
6/-3.7/-4.4
} {
\node[v] (u\i_1) at (\colU, \ytop) {};
\node[vlabel, right=2pt of u\i_1] {$u_\i^1$};
\node[val, right=18pt of u\i_1] {$0$};

\node at (\colU, {(\ytop+\ybot)/2}) {$\vdots$};

\node[v] (u\i_9) at (\colU, \ybot) {};
\node[vlabel, right=2pt of u\i_9] {$u_\i^9$};
\node[val, right=18pt of u\i_9] {$0$};
}

% =================================================================
% EDGES
% =================================================================

% --- Red/Coral Edges: c to d_i ---
\begin{scope}[red!65!black, opacity=0.8]
\foreach \i in {1,...,9} {
\draw (c) -- (d\i);
}
\end{scope}

% --- Green Edges: c to x_i ---
\begin{scope}[draw=green!60!black, thick]
\foreach \i in {1,...,6} {
\draw (c) -- (x\i);
}
\end{scope}

% --- Blue Edges: c to S_j ---
\begin{scope}[draw=blue!75!black, thick]
\foreach \j in {1,...,6} {
\draw (c) -- (s\j);
}
\end{scope}

% --- Yellow Edges: x_j to S_j ---
\begin{scope}[yellow!85!black, thick]
\foreach \j in {1,...,6} {
\draw (x\j) -- (s\j);
}
\end{scope}

% --- Orange Edges: Set-Element Incidence Edges (S_j to u_i^1, u_i^9) ---
\begin{scope}[orange!85!black, opacity=0.85]
% S1 = {1, 2, 3}
\foreach \elem in {1, 2, 3} {
\draw (s1) -- (u\elem_1);
\draw (s1) -- (u\elem_9);
}

% S2 = {1, 2, 4}
\foreach \elem in {1, 2, 4} {
\draw (s2) -- (u\elem_1);
\draw (s2) -- (u\elem_9);
}

% S3 = {1, 5, 6}
\foreach \elem in {1, 5, 6} {
\draw (s3) -- (u\elem_1);
\draw (s3) -- (u\elem_9);
}

% S4 = {3, 5, 6}
\foreach \elem in {3, 5, 6} {
\draw (s4) -- (u\elem_1);
\draw (s4) -- (u\elem_9);
}

% S5 = {2, 4, 5}
\foreach \elem in {2, 4, 5} {
\draw (s5) -- (u\elem_1);
\draw (s5) -- (u\elem_9);
}

% S6 = {3, 4, 6}
\foreach \elem in {3, 4, 6} {
\draw (s6) -- (u\elem_1);
\draw (s6) -- (u\elem_9);
}
\end{scope}

% --- Matching core edges from c' ---
\begin{scope}[red!65!black, opacity=0.8]
\foreach \i in {1,...,9} {
\draw (cp) -- (d\i);
}
\end{scope}

\begin{scope}[draw=green!60!black, thick]
\foreach \i in {1,...,6} {
\draw (cp) -- (x\i);
}
\end{scope}

\begin{scope}[draw=blue!75!black, thick]
\foreach \j in {1,...,6} {
\draw (cp) -- (s\j);
}
\end{scope}

\end{tikzpicture}
\end{adjustbox}
\caption{Size-two perfect Italian reduction for the $q=2$ RXC3 instance in Section~\ref{sec:prelim}, with exact cover $\{S_2,S_4\}$ and $r=9$. Red labels give a PIDF: $c,c'$ have value 1, $S_2,S_4$ have value 2, and $x_2,x_4$ have value 1; all other displayed vertices have value 0. Only copies $u_i^1,u_i^9$ are shown; vertical dots omit $u_i^2,\ldots,u_i^8$. Orange edges show set-element incidence. Both $c$ and $c'$ are adjacent to all set vertices, private tests, and forcing vertices, but not to element copies or to one another.}
\label{fig:pi-two-side}
\end{figure}

Given an exact cover, label $c,c'$ by 1, the selected set vertices
by 2, and their private tests $x_j$ by 1.  Label all remaining
vertices by zero.  Each zero set vertex, zero private test, and
$d_\ell$ receives one unit from each core vertex.  Each element
copy receives two units from its unique selected set vertex.
Hence this is a PIDF of weight $2+2q+q=K$.

Conversely, let $f$ be a PIDF of weight at most $K$.  At least one
of the $r>K$ vertices $d_\ell$ is zero, giving
\[
 f(c)+f(c')=2.
\]
Every element has a zero copy, so
$\sum_{j:e_i\in S_j}f(s_j)=2$ for every $i$.  Summing over the
$3q$ elements gives $\sum_j f(s_j)=2q$.  Let
$P=\{j:f(s_j)>0\}$.  For $j\in P$, the vertex $x_j$ cannot be
zero, since its neighbor sum would be $2+f(s_j)>2$.  Therefore
\[
 K\geq\wt(f)\geq2+2q+|P|,
 \qquad |P|\geq q.
\]
Equality follows: $|P|=q$, and every positive set vertex has label 2.
The element equations now say that these $q$ triples form an exact
cover.  Moreover, neither $c$ nor $c'$ can be zero: either would
receive at least $2q>2$ from $B$.  Thus their labels are indeed
$(1,1)$.  Membership in NP completes the proof.
\end{proof}

\begin{proof}[Theorem~\ref{thm:all-two-side}]
Lemma~\ref{lem:core-extension} with $t=2$ proves NP-hardness for Roman,
perfect Roman, and Italian domination; Lemma~\ref{lem:pi-two-side}
proves it for perfect Italian domination.  All four problems belong to
NP, so each is NP-complete.
\end{proof}

\subsection{Hardness with prescribed core labels for perfect Italian domination}
\label{sec:large-sides}
\label{subsec:pi-large-core}

Using the RXC3 framework from Section~\ref{sec:prelim}, we encode set
choices in vertices of $I$ and exact-cover constraints in the neighborhoods
of zero-labeled core vertices.  This yields NP-completeness of the
prescribed-core extension problem; the construction's low-weight rigidity
also gives the optimum gap used later for approximation hardness.

\begin{theorem}\label{cor:pi-prescribed-core}
Given a bipartite bisplit graph with a supplied partition and prescribed
labels on $A\cup B$, deciding whether these labels extend to a PIDF is
NP-complete, even when $|A|,|B|\geq3$ and the prescribed labeling assigns
2 to one vertex of $A$ and 0 to every other core vertex.  The same problem
with an additional weight bound is also NP-complete.
\end{theorem}

\paragraph{Construction.}
For an integer $r\geq1$, construct $J_r=J_r(U,\mathcal S)$ with partition
\[
\begin{aligned}
 A={}&\{p,z,h_{12},h_{13},h_{23}\}
       \mathbin{\dot\cup}\{w_j:1\leq j\leq 3q\}
       \mathbin{\dot\cup}\{e_i:1\leq i\leq3q\},\\
 B={}&\{b_\ell:1\leq\ell\leq r\},\\
 I={}&\{t_1,t_2,t_3,y\}
       \mathbin{\dot\cup}\{u_j,v_j:1\leq j\leq 3q\}
       \mathbin{\dot\cup}\{d_\ell:1\leq\ell\leq r\}.
\end{aligned}
\]
All three parts are stable.  Add every edge between $A$ and $B$, and join
$p$ to every vertex of $I\setminus\{y\}$.  The other edges are specified
by the following independent-set neighborhoods, where
$N_I(w)=N_{J_r}(w)\cap I$:
\begin{equation}\label{eq:large-core-incidences}
\begin{aligned}
 N_I(z)&=\{y\},\\
 N_I(h_{12})&=\{t_1,t_2\},\quad
 N_I(h_{13})=\{t_1,t_3\},\quad
 N_I(h_{23})=\{t_2,t_3\},\\
 N_I(w_j)&=\{u_j,v_j,t_1\}\quad(1\leq j\leq 3q),\\
 N_I(e_i)&=\{t_1\}\cup\{u_j:e_i\in S_j\}
                          \quad(1\leq i\leq3q).
\end{aligned}
\end{equation}
There are no other edges.  In particular, $y$ is a leaf of $z$, each
$d_\ell$ is a leaf of $p$, and there are no edges between $I$ and $B$.
Thus $(I,A,B)$ is a bisplit partition, and $(A,B\cup I)$ is a bipartition.
The part sizes are
\[
 |A|=6q+5,\qquad |B|=r,\qquad |I|=6q+r+4.
\]
The order and size of $J_r$ are $|V(J_r)|=12q+9+2r$ and
$|E(J_r)|=(6q+6)r+27q+10$, respectively. The construction is polynomial whenever
$r$ is polynomial in the source size.  Put $b=3q+7$.  For decision
hardness we take $r=b+1$; then both core sides grow with the instance.
Fig.~\ref{fig:pi-large-core-overview} gives the full overview;
Fig.~\ref{fig:pi-large-core-gadgets} shows the local equations.

\begin{reductionfigure}
\begin{adjustbox}{max width=\textwidth}
\begin{tikzpicture}[
    v/.style={circle, fill=black, inner sep=0pt, minimum size=3.5pt},
    vlabel/.style={font=\small, inner sep=2pt},
    val/.style={font=\small, color=red!80!black, inner sep=2pt},
    column_head/.style={font=\large\bfseries, align=center},
    column_sub/.style={font=\small, align=center}
]

  % --- Column Coordinates ---
  \def\colB{0}
  \def\colA{5}
  \def\colI{10}

  % --- Column Headers ---
  \node[column_head] at (\colB, 11) {$B$};
  \node[column_sub]  at (\colB, 10.5) {R zero vertices};
  \node[column_head] at (\colA, 11) {$A$};
  \node[column_sub]  at (\colA, 10.5) {Constraint side};
  \node[column_head] at (\colI, 11) {$I$};
  \node[column_sub]  at (\colI, 10.5) {Independent set};

  % =================================================================
  % COLUMN A (Middle: Constraint side)
  % =================================================================
  \node[v] (p) at (\colA, 9.8) {};
  \node[vlabel, above left=1pt of p] {$P$};
  \node[val, left=18pt of p] {$2$};

  \node[v] (z) at (\colA, 9.2) {};
  \node[vlabel, left=2pt of z] {$z$};
  \node[val, left=18pt of z] {$0$};

  \node[v] (h12) at (\colA, 8.6) {};
  \node[vlabel, left=2pt of h12] {$h_{12}$};
  \node[val, left=18pt of h12] {$0$};

  \node[v] (h13) at (\colA, 7.8) {};
  \node[vlabel, left=2pt of h13] {$h_{13}$};
  \node[val, left=18pt of h13] {$0$};

  \node[v] (h23) at (\colA, 7.0) {};
  \node[vlabel, left=2pt of h23] {$h_{23}$};
  \node[val, left=18pt of h23] {$0$};

  \node[v] (w1) at (\colA, 6.3) {};
  \node[vlabel, left=2pt of w1] {$w_1$};
  \node[val, left=18pt of w1] {$0$};

  \node[v] (w2) at (\colA, 5.7) {};
  \node[vlabel, left=2pt of w2] {$w_2$};
  \node[val, left=18pt of w2] {$0$};

  \node[v] (w3) at (\colA, 5.1) {};
  \node[vlabel, left=2pt of w3] {$w_3$};
  \node[val, left=18pt of w3] {$0$};

  \node[v] (w4) at (\colA, 4.5) {};
  \node[vlabel, left=2pt of w4] {$w_4$};
  \node[val, left=18pt of w4] {$0$};

  \node[v] (w5) at (\colA, 3.9) {};
  \node[vlabel, left=2pt of w5] {$w_5$};
  \node[val, left=18pt of w5] {$0$};

  \node[v] (w6) at (\colA, 3.3) {};
  \node[vlabel, left=2pt of w6] {$w_6$};
  \node[val, left=18pt of w6] {$0$};

  \node[v] (e1) at (\colA, 2.6) {};
  \node[vlabel, left=2pt of e1] {$e_1$};
  \node[val, left=18pt of e1] {$0$};

  \node[v] (e2) at (\colA, 2.0) {};
  \node[vlabel, left=2pt of e2] {$e_2$};
  \node[val, left=18pt of e2] {$0$};

  \node[v] (e3) at (\colA, 1.4) {};
  \node[vlabel, left=2pt of e3] {$e_3$};
  \node[val, left=18pt of e3] {$0$};

  \node[v] (e4) at (\colA, 0.8) {};
  \node[vlabel, left=2pt of e4] {$e_4$};
  \node[val, left=18pt of e4] {$0$};

  \node[v] (e5) at (\colA, 0.2) {};
  \node[vlabel, left=2pt of e5] {$e_5$};
  \node[val, left=18pt of e5] {$0$};

  \node[v] (e6) at (\colA, -0.4) {};
  \node[vlabel, left=2pt of e6] {$e_6$};
  \node[val, left=18pt of e6] {$0$};

  % =================================================================
  % COLUMN B (Left: R zero vertices)
  % =================================================================
  \node[v] (b1) at (\colB, 9.8) {};
  \node[vlabel, left=2pt of b1] {$b_1$};
  \node[val, left=18pt of b1] {$0$};

  \node[v] (b2) at (\colB, 9.1) {};
  \node[vlabel, left=2pt of b2] {$b_2$};
  \node[val, left=18pt of b2] {$0$};

  \node[v] (b3) at (\colB, 8.4) {};
  \node[vlabel, left=2pt of b3] {$b_3$};
  \node[val, left=18pt of b3] {$0$};

  \node at (\colB, 6.5) {$\cdot$};
  \node at (\colB, 4.6) {$\cdot$};
  \node at (\colB, 2.7) {$\cdot$};

  \node[v] (b14) at (\colB, 0.8) {};
  \node[vlabel, below left=1pt of b14] {$b_{14}$};
  \node[val, left=18pt of b14] {$0$};

  % =================================================================
  % COLUMN I (Right: Independent set)
  % =================================================================
  \node[v] (d1) at (\colI, 9.8) {};
  \node[vlabel, right=2pt of d1] {$d_1$};
  \node[val, right=18pt of d1] {$0$};

  \node[v] (d2) at (\colI, 9.3) {};
  \node[vlabel, right=2pt of d2] {$d_2$};
  \node[val, right=18pt of d2] {$0$};

  \node[v] (d3) at (\colI, 8.8) {};
  \node[vlabel, right=2pt of d3] {$d_3$};
  \node[val, right=18pt of d3] {$0$};

  \node[red] at (\colI-1.6, 8.95) {$\cdot$};
  \node[red] at (\colI-1.6, 8.85) {$\cdot$};
  \node[red] at (\colI-1.6, 8.75) {$\cdot$};
  \node at (\colI, 8.65) {$\cdot$};
  \node at (\colI, 8.5) {$\cdot$};
  \node at (\colI, 8.35) {$\cdot$};

  \node[v] (d14) at (\colI, 8.1) {};
  \node[vlabel, right=2pt of d14] {$d_{14}$};
  \node[val, right=18pt of d14] {$0$};

  \node[v] (y) at (\colI, 7.7) {};
  \node[vlabel, right=2pt of y] {$y$};
  \node[val, right=18pt of y] {$2$};

  \node[v] (t1) at (\colI, 7.0) {};
  \node[vlabel, right=2pt of t1] {$t_1$};
  \node[val, right=18pt of t1] {$1$};

  \node[v] (t2) at (\colI, 6.3) {};
  \node[vlabel, right=2pt of t2] {$t_2$};
  \node[val, right=18pt of t2] {$1$};

  \node[v] (t3) at (\colI, 5.6) {};
  \node[vlabel, right=2pt of t3] {$t_3$};
  \node[val, right=18pt of t3] {$1$};

  \node[v] (u1) at (\colI, 4.9) {};
  \node[vlabel, right=2pt of u1] {$u_1$};
  \node[val, right=18pt of u1] {$0$};

  \node[v] (u2) at (\colI, 4.3) {};
  \node[vlabel, right=2pt of u2] {$u_2$};
  \node[val, right=18pt of u2] {$1$};

  \node[v] (u3) at (\colI, 3.7) {};
  \node[vlabel, right=2pt of u3] {$u_3$};
  \node[val, right=18pt of u3] {$0$};

  \node[v] (u4) at (\colI, 3.1) {};
  \node[vlabel, right=2pt of u4] {$u_4$};
  \node[val, right=18pt of u4] {$1$};

  \node[v] (u5) at (\colI, 2.5) {};
  \node[vlabel, right=2pt of u5] {$u_5$};
  \node[val, right=18pt of u5] {$0$};

  \node[v] (u6) at (\colI, 1.9) {};
  \node[vlabel, right=2pt of u6] {$u_6$};
  \node[val, right=18pt of u6] {$0$};

  \node[v] (v1) at (\colI, 1.3) {};
  \node[vlabel, right=2pt of v1] {$v_1$};
  \node[val, right=18pt of v1] {$1$};

  \node[v] (v2) at (\colI, 0.7) {};
  \node[vlabel, right=2pt of v2] {$v_2$};
  \node[val, right=18pt of v2] {$0$};

  \node[v] (v3) at (\colI, 0.1) {};
  \node[vlabel, right=2pt of v3] {$v_3$};
  \node[val, right=18pt of v3] {$1$};

  \node[v] (v4) at (\colI, -0.5) {};
  \node[vlabel, right=2pt of v4] {$v_4$};
  \node[val, right=18pt of v4] {$0$};

  \node[v] (v5) at (\colI, -1.1) {};
  \node[vlabel, right=2pt of v5] {$v_5$};
  \node[val, right=18pt of v5] {$1$};

  \node[v] (v6) at (\colI, -1.7) {};
  \node[vlabel, right=2pt of v6] {$v_6$};
  \node[val, right=18pt of v6] {$1$};

  % =================================================================
  % EDGES
  % =================================================================

  % --- Blue Edges (Column B to Column A) ---
  \begin{scope}[cyan!65!blue, opacity=0.75]
    \foreach \dest in {p, z, h12, h13, h23, w1, w2, w3, w4, w5, w6, e1, e2, e3, e4, e5, e6} {
      \draw (b1) -- (\dest);
      \draw (b2) -- (\dest);
      \draw (b3) -- (\dest);
      \draw (b14) -- (\dest);
    }
  \end{scope}
  \node[blue] at (1.5, 4.3) {$\cdot$};
  \node[blue] at (1.5, 4.6) {$\cdot$};
  \node[blue] at (1.5, 4.9) {$\cdot$};

  % --- Red/Coral Edges (P to I) ---
  \begin{scope}[red!65!black, opacity=0.8]
    \draw (p) -- (d1);
    \draw (p) -- (d2);
    \draw (p) -- (d3);
    \draw (p) -- (d14);
    \foreach \dest in {t1, t2, t3, u1, u2, u3, u4, u5, u6, v1, v2, v3, v4, v5, v6} {
      \draw (p) -- (\dest);
    }
  \end{scope}

  % --- Green Edge ---
  \draw[green!60!black, thick] (z) -- (y);

  % --- Yellow Edges ---
  \begin{scope}[yellow!85!black, thick]
    \draw (h12) -- (t1); \draw (h12) -- (t2);
    \draw (h13) -- (t1); \draw (h13) -- (t3);
    \draw (h23) -- (t2); \draw (h23) -- (t3);
  \end{scope}

  % --- Purple Edges ---
  \begin{scope}[purple!80!black, thick]
    \foreach \i in {1,...,6} {
      \draw (w\i) -- (t1);
      \draw (w\i) -- (u\i);
      \draw (w\i) -- (v\i);
    }
  \end{scope}

  % --- Orange Edges ---
  \begin{scope}[orange!85!black, opacity=0.85]
    \draw (e1) -- (t1); \draw (e1) -- (u1); \draw (e1) -- (u2); \draw (e1) -- (u3);
    \draw (e2) -- (t1); \draw (e2) -- (u1); \draw (e2) -- (u2); \draw (e2) -- (u5);
    \draw (e3) -- (t1); \draw (e3) -- (u1); \draw (e3) -- (u4); \draw (e3) -- (u6);
    \draw (e4) -- (t1); \draw (e4) -- (u2); \draw (e4) -- (u5); \draw (e4) -- (u6);
    \draw (e5) -- (t1); \draw (e5) -- (u3); \draw (e5) -- (u4); \draw (e5) -- (u5);
    \draw (e6) -- (t1); \draw (e6) -- (u3); \draw (e6) -- (u4); \draw (e6) -- (u6);
  \end{scope}

\end{tikzpicture}
\end{adjustbox}
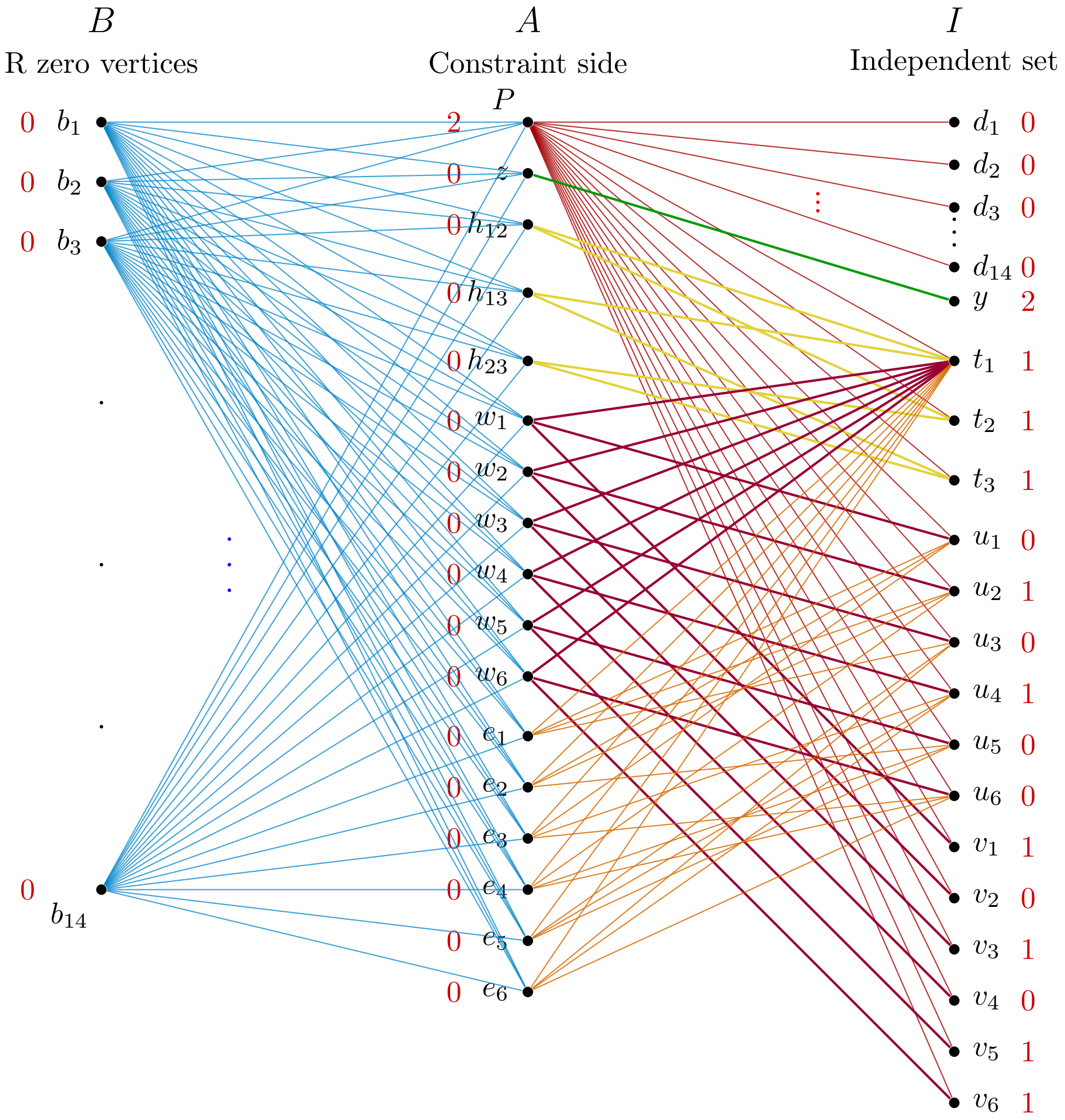
\caption{Illustration of the large-core perfect Italian reduction for
$q=2$ and $r=14$. Red numbers give the PIDF labels for the example
exact cover $\{S_2,S_4\}$ from Section~\ref{sec:prelim}. The sets $A$
and $B$ induce a complete bipartite graph.
Black dots between $b_3$ and $b_{14}$ indicate the omitted intermediate
vertices of $B$, and blue dots in the gap between $A$ and $B$ indicate
their omitted complete-bipartite edges to $A$. Dots in the $I$ column
indicate intermediate indexed vertices. The colored edges distinguish core edges and the neighborhoods used by
the gadgets. Orange edges show incidences from the common RXC3 instance
in Section~\ref{sec:prelim}.
Fig.~\ref{fig:pi-large-core-gadgets} details the local constraints.}
\label{fig:pi-large-core-overview}
\end{reductionfigure}

\begin{reductionfigure}
\begin{adjustbox}{max width=\textwidth}
\begin{tikzpicture}[
 v/.style={circle,fill=black,inner sep=0pt,minimum size=4pt},
 lab/.style={font=\small,inner sep=1pt},
 unitedge/.style={draw=yellow!75!black}, selector/.style={draw=purple}, cover/.style={draw=orange!80!black}]
 \foreach \name/\x in {t1/0,t2/1.1,t3/2.2,ta/3.7,u/4.8,v/5.9,tb/7.1,uone/8.2,utwo/9.3,uthree/10.4}
   \node[v] (\name) at (\x,1.2) {};
 \foreach \name/\x in {h12/0,h13/1.1,h23/2.2,w/4.8,e/8.75}
   \node[v] (\name) at (\x,-0.2) {};
 \draw[unitedge] (h12)--(t1) (h12)--(t2) (h13)--(t1) (h13)--(t3)
       (h23)--(t2) (h23)--(t3);
 \draw[selector] (w)--(ta) (w)--(u) (w)--(v);
 \draw[cover] (e)--(tb) (e)--(uone) (e)--(utwo) (e)--(uthree);
 \foreach \name/\text/\value in {t1/t_1/1,t2/t_2/1,t3/t_3/1,ta/t_1/1,u/u_j/1,v/v_j/0,tb/t_1/1,uone/u_{j_1}/1,utwo/u_{j_2}/0,uthree/u_{j_3}/0}{
   \node[lab,above=2pt of \name] {$\text$};
   \node[pidvalue,above=14pt of \name] {$\value$};
 }
 \foreach \name/\text in {h12/h_{12},h13/h_{13},h23/h_{23},w/w_j,e/e_i}{
   \node[lab,below=2pt of \name] {$\text$};
   \node[pidvalue,below=15pt of \name] {$0$};
 }
 \node[lab] at (1.1,2.3) {(a) unit labels};
 \node[lab] at (4.8,2.3) {(b) selector pair};
 \node[lab] at (8.75,2.3) {(c) exact cover};
\end{tikzpicture}
\end{adjustbox}
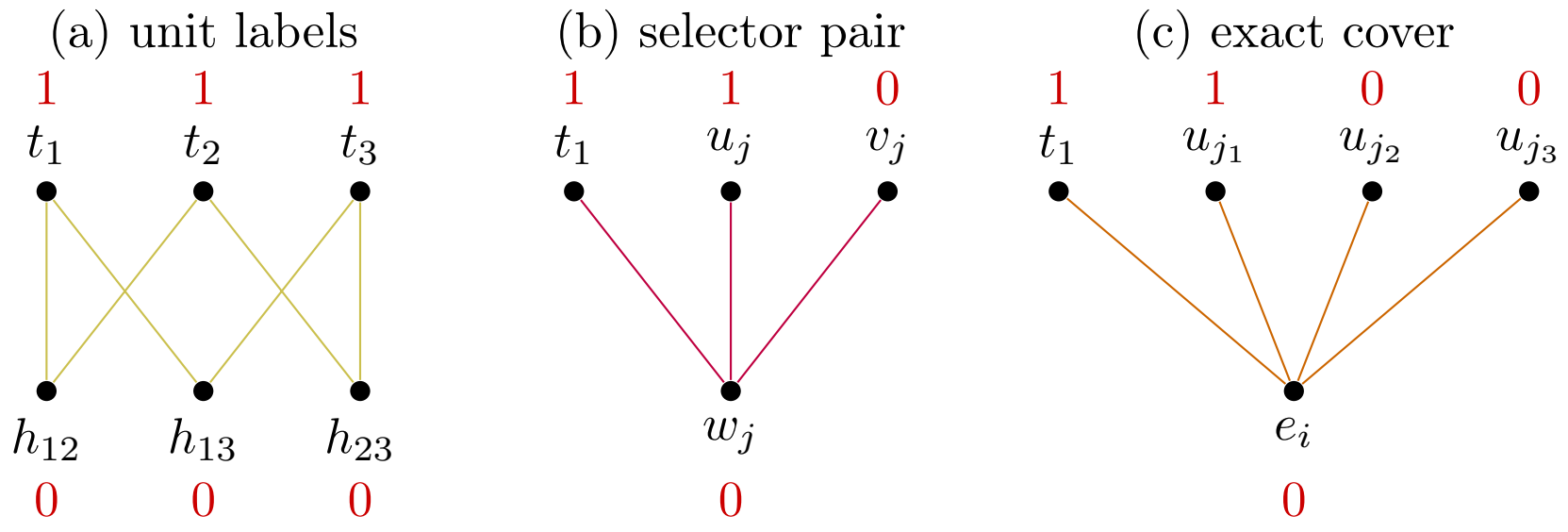
\caption{Local equation gadgets for the large-core perfect Italian
reduction in Fig.~\ref{fig:pi-large-core-overview}. Upper vertices lie in $I$ and
lower vertices in $A$; Panel (a) shows the unit
labels on $t_1,t_2,t_3$ and the pair constraints; (b) shows selected set
$S_j$ with $(f(u_j),f(v_j))=(1,0)$; (c) shows that among the three sets
containing $e_i$, exactly $S_{j_1}$ is selected.}
\label{fig:pi-large-core-gadgets}
\end{reductionfigure}

\begin{lemma}\label{lem:pi-large-core}
An exact cover gives a PIDF of $J_r$ of weight $b=3q+7$, for every
$r\geq1$.  Conversely, every PIDF of weight less than $r$ forces
$f(p)=2$, $f(A\setminus\{p\})=0$, and $f(B)=0$, and determines an
exact cover.  Thus, for $r>b$, a YES-instance has optimum $b$ and
a NO-instance has optimum at least $r$.
\end{lemma}

\begin{proof}
Suppose first that $\mathcal S'\subseteq\mathcal S$ is an exact cover.
Set
\[
 f(p)=f(y)=2,\qquad f(t_1)=f(t_2)=f(t_3)=1,
\]
and, for each $j$, put
\[
 (f(u_j),f(v_j))=
 \begin{cases}
 (1,0),&S_j\in\mathcal S',\\
 (0,1),&S_j\notin\mathcal S'.
 \end{cases}
\]
Assign zero to every other vertex.  Each zero vertex of $B$ receives two
from $p$.  Each zero vertex in $I$ also receives two from $p$, and its
other neighbors are zero constraint vertices in $A$.  The zero vertex
$z$ receives two from $y$, and each $h_{ij}$ receives one from each of
$t_i,t_j$.  Each $w_j$ receives one from $t_1$ and one from its pair
$u_j,v_j$.  Finally, each $e_i$ receives one from $t_1$ and one from the
unique selected set containing $e_i$.  Thus this is a PIDF, of weight
\[
 2+2+3+3q=3q+7=b.
\]

Conversely, let $f$ be a PIDF of weight less than $r$ and write
$\alpha=f(A)$, $\beta=f(B)$.  If $f(p)\ne2$, all $r$ private leaves
$d_\ell$ must be positive, giving weight at least $r$.  Therefore
$f(p)=2$.  At least one vertex of $B$ is zero because $|B|=r>\wt(f)$.
Since its neighborhood is exactly $A$, its constraint gives
$\alpha=2$.  Consequently, every vertex of $A\setminus\{p\}$ is zero.

The vertex $y$ has only the zero neighbor $z$, so $f(y)>0$.  The constraint
at $z$ is
\begin{equation}\label{eq:large-core-beta}
 \beta+f(y)=2,
\end{equation}
which implies $\beta\leq1$.  The three zero vertices $h_{ij}$ give
\[
 f(t_1)+f(t_2)=f(t_1)+f(t_3)=f(t_2)+f(t_3)=2-\beta.
\]
Subtracting these equations shows that all three labels are equal to
$(2-\beta)/2$.  Since labels are integral and $\beta\in\{0,1\}$, we
obtain
\[
 \beta=0,\qquad f(t_1)=f(t_2)=f(t_3)=1,
 \qquad f(y)=2.
\]
Now each zero vertex $w_j$ imposes
\[
 f(u_j)+f(v_j)+1=2.
\]
Hence $(f(u_j),f(v_j))$ is either $(1,0)$ or $(0,1)$.  The constraint at
each zero element vertex $e_i$ is
\[
 1+\sum_{j:e_i\in S_j}f(u_j)=2.
\]
Therefore every element belongs to exactly one member of
$\{S_j:f(u_j)=1\}$.  These sets form an exact cover; as each has size
three, there are exactly $q$ of them.  The forced labels have total
weight $b=3q+7$, proving both the rigidity and the claimed optimum gap.
\end{proof}

\begin{proof}[Theorem~\ref{cor:pi-prescribed-core}]
Take $J_{b+1}$ and prescribe $f(p)=2$, with all other core values zero.
In \emph{any} extension, the zero vertex $z$ forces $f(y)=2$; the
three pair equations at $h_{12},h_{13},h_{23}$ force $f(t_i)=1$.
The equations at $w_j$ then give $f(u_j)+f(v_j)=1$, and those at
$e_i$ give $\sum_{j:e_i\in S_j}f(u_j)=1$.  Hence the value-1
selectors specify an exact cover, regardless of the total weight.
Conversely, an exact cover supplies the extension of weight $b$
in Lemma~\ref{lem:pi-large-core}.  This proves both the budget-free
claim and the version with bound $b$.  A full labeling is a
polynomially checkable certificate in either case.
\end{proof}

Thus, even with these core labels prescribed, deciding whether they extend
to a PIDF remains NP-complete: the labels leave the exact-cover choices
encoded in $I$.

\begin{theorem}\label{thm:both-sides}
The decision problems for Roman, perfect Roman, Italian, and perfect Italian
domination are NP-complete on bisplit graphs supplied with a partition
satisfying $|A|,|B|\geq3$.
\end{theorem}

\begin{proof}
For Roman, perfect Roman, and Italian domination,
Lemma~\ref{lem:core-extension} gives NP-hardness on the subclass
with $|A|, |B|\geq3$.  For perfect Italian domination,
Lemma~\ref{lem:pi-large-core} with $r=b+1$ and budget $b$ gives
NP-hardness; its construction has $|A|=6q+5$ and $|B|=3q+8$.
All four problems belong to NP by Section~\ref{subsec:graph-prelim}, so
they are NP-complete.
\end{proof}

We next derive approximation bounds, which require more than the decision
equivalences alone.

\section{Approximation Hardness}\label{sec:approx}

Set Cover supplies the nonperfect bounds; replication creates
exact-cover gaps for the perfect rules.

\subsection{Roman and Italian domination: Set Cover}
\label{subsec:cover-approx}

For comparison, Italian domination has a
$2(1+\ln(\Delta+1))$-approximation on general graphs, where $\Delta$
is the maximum degree~\citep[Theorem~15]{padamutham-r2}.  Thus our
Italian lower bound retains the logarithmic dependence on graph order
even under the bisplit restriction, up to constant factors.

Let $U=\{e_1,\ldots,e_p\}$ be nonempty and let
$\mathcal S=\{S_1,\ldots,S_m\}$ cover $U$.  Here sets have arbitrary
size, unlike the triples in the decision sections.  Write $\tau$ for
the minimum cover size.  Fix $t\in\{1,2,3\}$ and put
$r=(p+m+1)^3$.  Construct $G_t$ with
\[
\begin{aligned}
 A&=\{c,a_1,\ldots,a_{t-1}\},\qquad B=\{s_1,\ldots,s_m\},\\
 I&=\{d_\ell:1\leq\ell\leq r\}
 \mathbin{\dot\cup}\{u_i^\ell:1\leq i\leq p,\ 1\leq\ell\leq r\}.
\end{aligned}
\]
For $t=1$, $A=\{c\}$.  Add all $A$--$B$ edges, let $d_\ell$ be a
private leaf of $c$, and join $u_i^\ell$ to $s_j$ exactly when
$e_i\in S_j$.  There are no other edges; in particular, this graph
has no private set tests $x_j$.  Its order is
$n=t+m+(p+1)r$.  Fig.~\ref{fig:roman-construction} shows $G_1$.
The extra core vertices for $t=2,3$ have label zero and are adjacent
exactly to $B$.

\begin{reductionfigure}
\begin{adjustbox}{max width=\textwidth}
\begin{tikzpicture}[
 v/.style={circle,draw=black,fill=black,minimum size=5pt,inner sep=0pt},
 vlabel/.style={font=\small,inner sep=1pt},
 lab/.style={font=\small,align=center},
 every edge/.style={draw}]
 \node[v] (c) at (0,0) {};
 \node[vlabel,below=2pt of c] {$c$};
 \node[lab,above=3mm of c] {$A=\{c\}$};

 \node[v] (s1) at (3.1,1.6) {};
 \node[v] (s2) at (3.1,0) {};
 \node[v] (sm) at (3.1,-1.6) {};
 \node[vlabel,above right=1pt and 1pt of s1] {$s_1$};
 \node[vlabel,above=2pt of s2] {$s_2$};
 \node[vlabel,below=2pt of sm] {$s_m$};
 \node at (3.1,-0.8) {$\vdots$};
 \node[lab,above=3mm of s1] {$B$ (set vertices)};
 \draw[blue!55,thick] (c)--(s1) (c)--(s2) (c)--(sm);

 \node[v] (d1) at (-3.1,1.1) {};
 \node[v] (dr) at (-3.1,-1.1) {};
 \node[vlabel,left=2pt of d1] {$d_1$};
 \node[vlabel,left=2pt of dr] {$d_r$};
 \node at (-3.1,0) {$\vdots$};
 \node[lab,above=3mm of d1] {forcing vertices};
 \draw (c)--(d1) (c)--(dr);

 \node[v] (u11) at (6.3,1.9) {};
 \node[v] (u1r) at (6.3,0.9) {};
 \node[v] (up1) at (6.3,-0.9) {};
 \node[v] (upr) at (6.3,-1.9) {};
 \node[vlabel,right=2pt of u11] {$u_1^1$};
 \node[vlabel,right=2pt of u1r] {$u_1^r$};
 \node[vlabel,right=2pt of up1] {$u_p^1$};
 \node[vlabel,right=2pt of upr] {$u_p^r$};
 \node at (6.3,1.4) {$\vdots$};
 \node at (6.3,0) {$\vdots$};
 \node at (6.3,-1.4) {$\vdots$};
 \node[lab,above=3mm of u11] {$I$ (element copies)};
 \draw (s1)--(u11) (s1)--(u1r);
 \draw (s2)--(u11) (s2)--(u1r) (s2)--(up1) (s2)--(upr);
 \draw (sm)--(up1) (sm)--(upr);

 \node[pidvalue,below=15pt of c] {$2$};
 \node[pidvalue,left=18pt of d1] {$0$};
 \node[pidvalue,left=18pt of dr] {$0$};
 \node[pidvalue,above=2pt of s1,xshift=-9pt] {$0$};
 \node[pidvalue,above=16pt of s2] {$2$};
 \node[pidvalue,below=2pt of sm,xshift=16pt] {$0$};
 \foreach \u in {u11,u1r,up1,upr}
   \node[pidvalue,right=25pt of \u] {$0$};

 \node[lab,text width=10.5cm] at (1.6,-2.85)
   {$s_j u_i^\ell\in E(G)$ exactly when $e_i\in S_j$,
    $1\leq i\leq p$, $1\leq j\leq m$, $1\leq\ell\leq r$};
\end{tikzpicture}
\end{adjustbox}
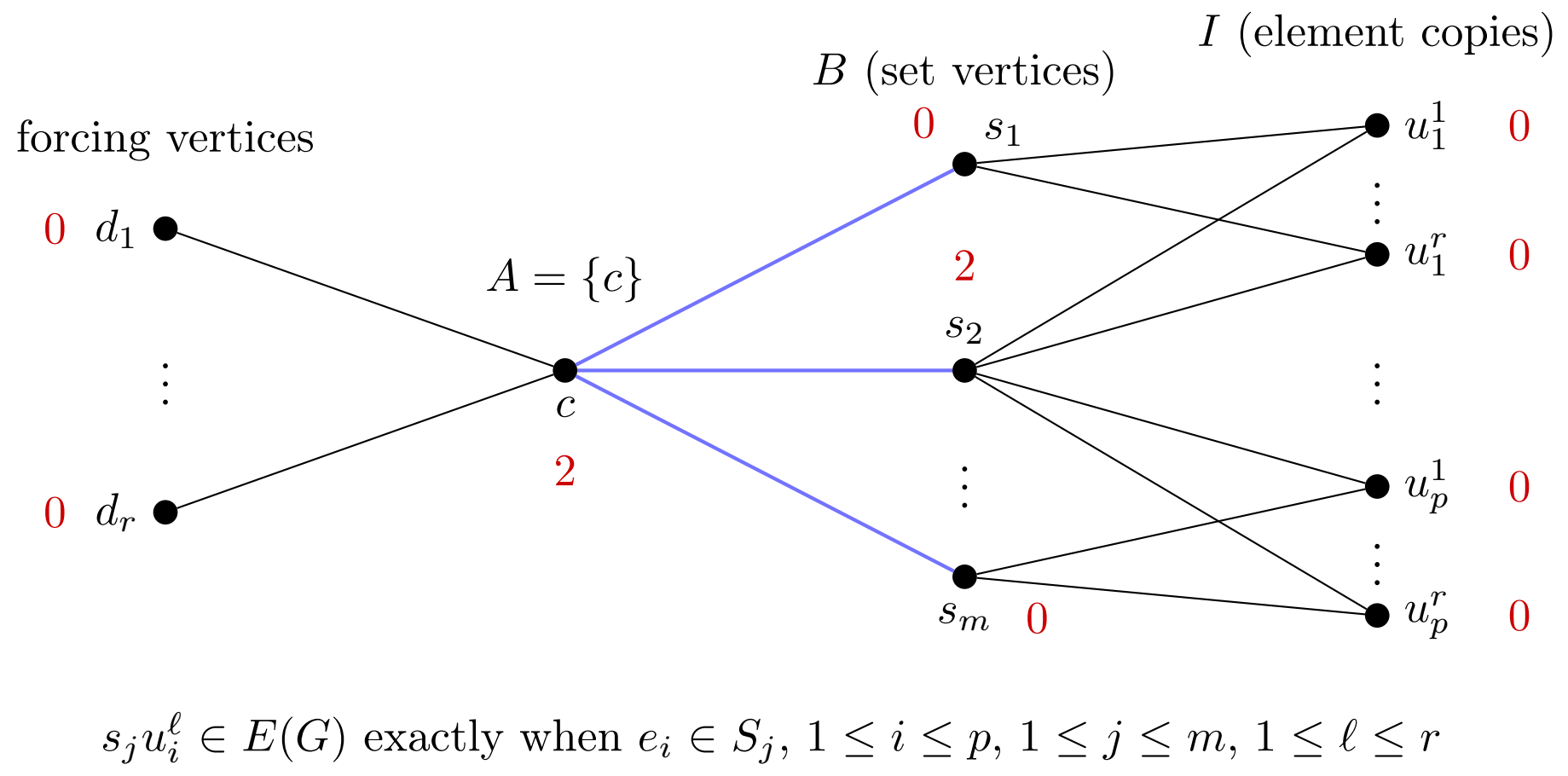
\caption{The Set Cover construction $G_1$ for Roman and Italian approximation hardness.
The blue edges join $c$ to all set vertices.  Each forcing vertex is adjacent
only to $c$, and each copy $u_i^\ell$ is adjacent exactly to the set vertices
whose sets contain $e_i$.  Only representative incidence edges are shown.
Red values show a labeling that is both an RDF and an IDF, obtained from a set cover, with $s_2$ selected and
$s_1,s_m$ unselected among the displayed set vertices.  The center and
selected set vertices have value 2; all other vertices have value 0.
Additional selected neighbors of an element copy are allowed}
\label{fig:roman-construction}
\end{reductionfigure}

\begin{theorem}\label{thm:cover-inapprox}
Unless $\mathrm{P}=\mathrm{NP}$, neither Roman domination nor Italian
domination admits a polynomial-time $o(\log n)$-approximation on bisplit
graphs supplied with a partition satisfying $|A|=t$ and $|B|\geq3$, for
any fixed $t\in\{1,2,3\}$.
\end{theorem}

\begin{proof}
A cover of size $\tau$ gives both an RDF and an IDF: label $c$ and
the selected set vertices 2, and everything else 0.  Its weight is
$2+2\tau$; any added $a_i$ sees a selected set.  Selecting all sets
costs at most $2+2m<r$, so both optimum weights are below $r$.

Any RDF or IDF of weight $w<r$ has $f(c)=2$ and a zero copy of each
element.  For Roman domination, the value-2 set vertices therefore
form a cover of size at most $(w-2)/2$.  For Italian domination,
the positive set vertices form a cover of size at most $w-2$.
In particular,
\[
 \gamma_{\R}(G_t)=2+2\tau,\qquad
 2+\tau\leq\gamma_{\Ita}(G_t)\leq2+2\tau.
\]
The Italian extraction allows two value-1 neighbors; it does not
use the invalid inference discussed in Section~\ref{subsec:italian-singleton}.

Suppose either problem has approximation ratio
$\rho(n)=o(\log n)$.  For sufficiently large inputs, the returned
weight is at most $\rho(n)(2+2m)<r$.  The extracted cover has size
at most $2\rho(n)\tau$ for Roman domination and $4\rho(n)\tau$ for
Italian domination, since $\tau\geq1$.  On the hard Set Cover family
underlying the logarithmic lower bound, $m$ is polynomially bounded
in $p$~\citep{dinur-steurer}.  Thus $\log n=O(\log p)$, and either
algorithm would give an $o(\log p)$-approximation for Set Cover,
a contradiction.  Instances with $m<3$ or bounded size can be
handled directly.
\end{proof}

\subsection{The perfect variants: an exact-cover gap}

The perfect Italian factor below is known for general graphs
\citep{pradhan-pi-complexity}.  Our proof retains it under the additional
bipartite bisplit restriction.  It does not follow merely by applying
the general-graph result to a subclass.

\begin{theorem}\label{thm:perfect-inapprox}
Fix $0<\varepsilon<1$.  Unless $\mathrm{P}=\mathrm{NP}$, there is no
polynomial-time $n^{1-\varepsilon}$-approximation for either of the
following problems:
\begin{enumerate}
 \item perfect Roman domination on bisplit graphs with a supplied
 partition satisfying $|A|=t$, $|B|\geq3$, for any fixed
 $t\in\{1,2,3\}$;
 \item perfect Italian domination on bipartite bisplit graphs with
 a supplied partition whose two core sides grow with the input.
\end{enumerate}
\end{theorem}

\begin{proof}
Use the common RXC3 instance.  Put $N=4q+1$, choose an integer
$d>2/\varepsilon$, and set $L=N^d$.

\emph{Perfect Roman.} Set the replication parameter to $r=L$ in the
construction $H_r$ of Section~\ref{subsec:common-construction}, then add
$t-1$ vertices adjacent exactly to $B$ as in
Lemma~\ref{lem:core-extension}.
An exact cover gives a PRDF of weight
$b_{\PR}=2+3q+(t-1)$, with $f(c)=2$ and all added core labels 1.
Conversely, any PRDF of weight below $L$ has $f(c)=2$ and a zero
copy of every element.  These copies force the value-2 set vertices
to form an exact cover.  Hence a NO-instance has optimum at least
$L$.  The order is $n=t+6q+(3q+1)L$.

\emph{Perfect Italian.} Set the replication parameter to $r=L$ in the
construction $J_r$ of Section~\ref{subsec:pi-large-core}.
Lemma~\ref{lem:pi-large-core} gives a YES optimum $b_{\PI}=3q+7$ and a
NO optimum at least $L$.
Here $n=12q+9+2L$; the graph is bipartite and both core sides grow.

In both cases, a YES-instance has optimum at most $b=O(N)$,
a NO-instance has optimum at least $L=N^d$, and
$n=O(N^{d+1})$.  A purported approximation algorithm would return,
on every sufficiently large YES-instance, a feasible labeling of weight
\[
 n^{1-\varepsilon}b
 =O\bigl(N^{(d+1)(1-\varepsilon)+1}\bigr)<N^d=L,
\]
where the strict inequality follows from $\varepsilon(d+1)>2$.
No such labeling exists on a NO-instance.  Comparing the returned
weight with $L$ would therefore decide RXC3 in polynomial time;
bounded-size instances can be solved directly.
\end{proof}

The resulting lower bounds are logarithmic for the nonperfect rules and
$n^{1-\varepsilon}$ for the perfect rules; the latter Italian bound uses
the large core only.  We next bound the total core size, which is
unbounded in every hard family above, and give FPT algorithms for
graphs close to a small vertex cover, a cluster graph, or a union of
bicliques.

\section{Parameterized Results}\label{sec:parameterized}

\subsection{Parameterization by vertex-cover number}\label{subsec:vc-parameter}

The following elementary dynamic program works for any supplied vertex
cover, without requiring complete bipartite edges inside the core.
Indeed, for a vertex cover $C$, the bags $C\cup\{v\}$, $v\notin C$, give a tree decomposition
of width at most $|C|$ (use one bag $C$ if $C=V(G)$).  The point here
is the explicit state count, not new fixed-parameter tractability
beyond results such as \citep[Theorem~15]{mann-fernau-pr}.

\begin{theorem}\label{thm:fixed-core}
Given a graph $G$ with a supplied vertex cover $C$ of size $k\geq1$,
the optimum Roman, perfect Roman, Italian, and perfect Italian domination
weights can be computed, respectively, in $O(4^kkn)$, $O(5^kkn)$,
$O(5^kkn)$, and $O(6^kkn)$ time.
In particular, all four problems are polynomial-time solvable on bisplit
graphs with every fixed pair $(|A|,|B|)=(a,b)$.
\end{theorem}

\begin{proof}
Put $I=V(G)\setminus C$, which is stable.  Fix one rule and enumerate
the $3^k$ labelings $g:C\to\{0,1,2\}$.  For each $v\in I$, whether
label zero is allowed is determined entirely by $g$ on $N(v)\subseteq C$.
Let $L_g(v)$ contain 1 and 2, and also 0 precisely when this condition
holds.  Thus any choice from these lists satisfies all constraints in $I$.

Only the zero core vertices $Z=g^{-1}(0)$ still impose constraints.
For the Roman rules, define the contribution of a label $a$ as
$\delta(a)=\mathbf{1}_{\{a=2\}}$; for the Italian rules, put
$\delta(a)=a$.  Set the saturation threshold $h$ to 1, 2, 2, and 3
for Roman, perfect Roman, Italian, and perfect Italian domination,
respectively.  A state records each zero core vertex's accumulated
contribution in $\{0,\ldots,h\}$, where $h$ means \emph{at least $h$}.
Initialize the vector by
\[
 s_z^0=\min\!\left\{h,\sum_{u\in N(z)\cap C}\delta(g(u))\right\}
 \qquad(z\in Z).
\]
Process the vertices $v_1,\ldots,v_{|I|}$ in any order.  A table
$D_i(s)$ stores the minimum weight on the first $i$ vertices producing
state $s$.  Initially $D_0(s^0)=0$, with every other entry infinite.
Initialize each new table to infinity.  For each finite entry and
each $a\in L_g(v_{i+1})$, update
\[
 s'_z=\min\{h,s_z+\delta(a)\mathbf{1}_{\{zv_{i+1}\in E(G)\}}\},
 \qquad
 D_{i+1}(s')=\min\{D_{i+1}(s'),D_i(s)+a\}.
\]
At the end, every coordinate must be at least 1 for Roman, exactly 1
for perfect Roman, at least 2 for Italian, and exactly 2 for perfect
Italian domination.  The chosen saturation preserves each test,
including rejection of excessive contributions for the perfect rules.
Add $g(C)$ to the minimum accepted entry and minimize over $g$.

Every feasible labeling follows these transitions.  Conversely, every
accepted path satisfies the constraints in $I$ by the lists and those
in $C$ by the final state.  Taking minimum weights therefore gives the
optimum.  For a fixed $g$ there are at most $(h+1)^{|Z|}$ states, three
choices per transition, and $O(k)$ work per choice.  Summing the state
bound over all core assignments gives
$\sum_{g:C\to\{0,1,2\}}(h+1)^{|g^{-1}(0)|}=(h+3)^k$,
so the total time is $O((h+3)^kkn)$.  For a bisplit partition, take
$C=A\cup B$ and $k=a+b$.
\end{proof}

Thus every fixed pair $(a,b)$ is tractable, unlike bounding one side or
prescribing core labels.

\subsection{Parameterization by cluster vertex deletion}\label{subsec:cvd}

The cluster vertex deletion number, denoted $\operatorname{cvd}(G)$,
is the minimum size of a set $X\subseteq V(G)$ such that $G-X$ is a
disjoint union of cliques.  The Roman domination problem is already
known to be FPT for this parameter~\citep{ashok-rdf-cvd}.  The following
uniform dynamic program handles all four variants.

\begin{theorem}\label{thm:cvd}
Given a graph $G$ and a supplied set $X$ of size $k$ such that $G-X$ is
a disjoint union of cliques, the optimum Roman, perfect Roman, Italian,
and perfect Italian domination weights can be computed, respectively, in
$O(4^k n^{O(1)})$, $O(11^k n^{O(1)})$, $O(11^k n^{O(1)})$, and
$O(18^k n^{O(1)})$ time.  The Roman bound uses the known
$O(4^k n^{O(1)})$ algorithm~\citep{ashok-rdf-cvd}; the dynamic program
below gives the other three bounds.  Consequently, all four problems are
FPT parameterized by $\operatorname{cvd}(G)$.
\end{theorem}

\begin{proof}
Fix one of the four rules and a labeling $g:X\to\{0,1,2\}$.  For the
Roman rules, let $\delta(a)=\mathbf{1}_{\{a=2\}}$; for the Italian
rules, let $\delta(a)=a$.  Use the thresholds $h=1,2,2,3$ for Roman,
perfect Roman, Italian, and perfect Italian domination, respectively;
values are saturated at $h$.  Write $Z=g^{-1}(0)$.  For each clique
component $Q$ of $G-X$, enumerate its capped total contribution
$t\in\{0,\ldots,h\}$.  If a vertex $v\in Q$ has label zero, then
$\delta(0)=0$, so its contribution from the other vertices of $Q$ is
exactly this total.  Its contribution from $X$ is fixed by $g$.
Therefore, for each choice of $t$, we know whether label zero is
permitted at every vertex of $Q$.

For each $t$, a dynamic program processes the vertices of $Q$, choosing
among the permitted labels and recording the capped total contribution
on $Q$, the labeling weight, and a vector of capped contributions to
the vertices in $Z$.  Retain only states whose total on $Q$ is $t$;
these give the minimum cost for each contribution vector from $Q$ to
$Z$.  A second dynamic program combines these tables over the clique
components.  It starts with the contributions to $Z$ from $g$, adds
each component's vector with saturation at $h$, and accepts exactly
those final states satisfying the rule at every vertex of $Z$.  Adding
$g(X)$ and minimizing over all labelings $g$ gives the optimum.  The
tables encode every feasible labeling, and each accepted state is
feasible, so the algorithm is exact.

For a fixed $g$, the state vectors have at most $(h+1)^{|Z|}$ entries;
combining two tables takes $(h+1)^{2|Z|}n^{O(1)}$ time.  Summing over
all labelings of $X$ gives
\[
 \sum_{g:X\to\{0,1,2\}}(h+1)^{2|g^{-1}(0)|}
   =\bigl(2+(h+1)^2\bigr)^k.
\]
Thus this dynamic program takes $O(6^k n^{O(1)})$ time for Roman
domination, $O(11^k n^{O(1)})$ time for perfect Roman and Italian
domination, and $O(18^k n^{O(1)})$ time for perfect Italian domination.
For Roman domination we use the stronger known bound stated in the
theorem.  If the modulator is not supplied, it can be found by branching
on an induced $P_3$: cluster graphs are exactly the $P_3$-free graphs,
and every deletion set must hit each such obstruction.  This branching
has depth at most $k$ and at most three branches at each step, so it is
FPT as well.
\end{proof}

\subsection{Parameterization by biclique-cluster vertex deletion}\label{subsec:bcvd}

For a graph $G$, let $\operatorname{bcvd}(G)$ be the minimum size of a
set $X\subseteq V(G)$ such that every component of $G-X$ is a complete
bipartite graph; isolated vertices are allowed as components.  We give
an FPT algorithm for each of the four domination rules with respect to
this parameter.

\begin{theorem}\label{thm:bcvd}
Given a graph $G$ and a supplied set $X$ of size $k$ such that $G-X$ is a
disjoint union of complete bipartite graphs, the optimum Roman, perfect
Roman, Italian, and perfect Italian domination weights can be computed,
respectively, in $O(6^k n^{O(1)})$, $O(11^k n^{O(1)})$,
$O(11^k n^{O(1)})$, and $O(18^k n^{O(1)})$ time.  Consequently, all
four problems are fixed-parameter tractable parameterized by
$\operatorname{bcvd}(G)$.
\end{theorem}

\begin{proof}
Fix one of the four rules and enumerate the $3^k$ labelings
$g:X\to\{0,1,2\}$.  For Roman and perfect Roman domination, let
$\delta(a)=\mathbf{1}_{\{a=2\}}$; for Italian and perfect Italian
domination, let $\delta(a)=a$.  Use saturation thresholds
$h=1,2,2,3$, respectively, where the value $h$ records a contribution
of at least $h$.  Thus the threshold distinguishes satisfaction from
failure for the lower-bound rules and also detects excess for the exact
rules.  Only the zero-labeled vertices $Z=g^{-1}(0)$ of $X$ still need
to be checked after all components are processed.

Consider a nontrivial complete bipartite component $H$ with sides $L$
and $R$.  For each pair of capped totals
$(t_L,t_R)\in\{0,\ldots,h\}^2$, prescribe that the labels on each side
have total $\delta$-contribution $t_L$ or $t_R$, respectively.  For a
vertex $v\in L$ labeled zero, its contribution from $R$ is the sum of
the $\delta$-values on $R$, represented by $t_R$ after saturation,
since all $L$--$R$ edges are present; its contribution from $X$ is fixed
by $g$.  Hence, for this pair of totals, whether zero
is allowed at each vertex of $L$ is known.  The same holds for $R$.
On each side, a dynamic program chooses allowed labels while recording
the capped total on that side, the cost, and a vector of capped
contributions to the vertices of $Z$.  Pairing the two side tables
therefore gives, for every contribution vector to $Z$, the minimum
weight of a labeling of $H$ that satisfies all zero-vertex constraints
inside $H$.  Isolated vertices are handled directly using their fixed
neighbors in $X$.

Finally, a second dynamic program combines these component tables.  It
starts with the contributions to $Z$ from the labeling $g$ on $X$,
adds each component's contribution vector with saturation at $h$, and
accepts precisely the final vectors satisfying the applicable Roman,
perfect Roman, Italian, or perfect Italian condition at every vertex
of $Z$.  Adding $g(X)$ and minimizing over $g$ yields the optimum.
Every feasible labeling appears in these tables, and every accepted
combination satisfies all constraints, so the computed value is exact.

For fixed $g$, the contribution vectors have at most $(h+1)^{|Z|}$
entries; combining two such tables takes at most
$(h+1)^{2|Z|}n^{O(1)}$ time.  Summing over all labelings of $X$ gives
\[
 \sum_{g:X\to\{0,1,2\}}(h+1)^{2|g^{-1}(0)|}
   =\bigl(2+(h+1)^2\bigr)^k.
\]
since $h\leq3$.  Substituting $h=1,2,2,3$ gives the respective
running-time bases $6,11,11,18$ for Roman, perfect Roman, Italian, and
perfect Italian domination.  This proves the claimed running times when
a modulator is supplied.  To obtain an algorithm parameterized only by
$\operatorname{bcvd}(G)$, one can find such a modulator by branching on
an induced $P_4$ or triangle: a graph is a disjoint union of complete
bipartite graphs exactly when it has neither, and every valid deletion
set must hit each such obstruction.  The branching depth is at most
$k$ and the branching factor at most four, so this preprocessing is
also fixed-parameter tractable.
\end{proof}

We conclude by summarizing these results and discussing directions
for further research.

\section{Conclusions and Further Research}

We studied Roman, perfect Roman, Italian, and perfect Italian domination
on bisplit graphs.  Our reductions give NP-completeness proofs for all
four problems when one core side has size one or two, and when both
sides have size at least three.  We also identified a flaw in an earlier
Italian domination reduction and gave an independent proof of its
hardness claim.  For perfect Italian domination, fixing the core labels
does not make the extension problem easy: it remains NP-complete even
on bipartite bisplit graphs when the prescribed core weight is two and
there is no bound on the total weight.  The approximation results rule
out polynomial-time $o(\log n)$-approximations for Roman and Italian
domination and $n^{1-\varepsilon}$-approximations for the perfect variants
on the graph classes specified above, for every fixed
$0<\varepsilon<1$, unless $\mathrm{P}=\mathrm{NP}$.  In contrast, our
vertex-cover dynamic program computes the four optimum weights in
$O(c^kkn)$ time, with $c=4,5,5,6$, respectively.  Thus bounding the
total core size makes all four problems tractable, whereas bounding
only one side does not.  With supplied modulators of size $k$, the
running-time bases, in the order Roman, perfect Roman, Italian, and
perfect Italian, are $4,11,11,18$ for cluster vertex deletion and
$6,11,11,18$ for biclique-cluster vertex deletion, up to polynomial
factors.  The $4^k$ cluster-deletion bound for Roman domination is known;
the other bounds follow from our dynamic programs.  All four problems
are FPT for both parameters.

A question left open by this work is whether similar approximation
lower bounds hold for perfect Italian domination when one core side
has size one or two.  Another direction is to study restrictions on
the neighborhoods of vertices outside the core that allow efficient
algorithms, even when the core is large.  It would also be useful to
investigate whether the running times of FPT algorithms can be improved.

\section*{Statements and Declarations}

\paragraph{Funding.}
This research did not receive any specific grant from funding agencies in the public, commercial, or not-for-profit sectors.

\paragraph{Declaration of competing interests.}
The authors have nothing to declare regarding competing interests.

\paragraph{Data availability.}
No new empirical data were created or analyzed in this study.

\section*{Declaration of generative AI and AI-assisted technologies in the manuscript preparation process}
During the preparation of this work, the authors used OpenAI Prism to assist
with language editing, LaTeX formatting, and consistency checking. After using
this tool, the authors reviewed and edited the content as needed, verified
the correctness of the results, and take full responsibility for the content
of the published article.

\end{document}